\documentclass{aa}
\usepackage{graphicx}
\usepackage{natbib}
\bibpunct{(}{)}{;}{a}{}{,}
\begin{document}

\title{The different progenitors of type Ib, Ic SNe, and of GRB}
\titlerunning{Progenitors of type Ib, Ic SNe}

\author{Cyril Georgy$^1$, Georges Meynet$^1$, Rolf Walder$^{1,2}$, Doris Folini$^2$, Andr\'e Maeder$^1$}
\authorrunning{Georgy et al.}

\institute{$^1$Geneva Observatory, Geneva University, CH--1290 Sauverny, Switzerland\\
$^2$Ecole Normale Sup\'erieure, Lyon, CRAL (UMR CNRS 5574), Universit\'e de Lyon, France\\
email: cyril.georgy@obs.unige.ch}

\date{Received  / Accepted }

\offprints{Cyril Georgy} 
   
\abstract{}{We discuss the properties of the progenitors of core collapse supernovae of type Ib and Ic and of long soft gamma ray bursts, as they can be deduced from rotating stellar models of single stars at various metallicities.}{The type of the supernova progenitor was determined from the surface abundances at the pre-supernova stage. The type of the supernova event was obtained from the masses of hydrogen and helium ejected at the time of the core-collapse supernova event.}{We find that the minimum amount of helium ejected by a core-collapse supernova (of whatever type) is around 0.3 M$_\odot$. There is no difference between the WC and WO stars in the ejected masses of  helium, CNO elements, and heavy elements. Also no difference is expected between the chemical composition of a WC star resulting from a normal or a homogeneous evolution. The progenitors of type Ib supernovae are WNL, WNE, or less massive WC stars. Those of type Ic are WC and WO stars.  WO stars are produced in a limited mass range (around 60 M$_\odot$) and only at low metallicity (for $Z\le\sim 0.010$) as already found. The WO stars are the progenitors of only a small fraction of type Ic. Present stellar models indicate that, at solar metallicity, there is about 1 type Ib supernova for 1 type Ic, and this ratio rises to 3 type Ic for 1 type Ib SN at twice solar metallicity. At this metallicity, type Ic's are more frequent than type Ib's because most massive stars that go through a WNE stage evolve further into a WC/WO phase. Current models can account for the observed number ratios SN Ib/SN II and SN Ic/SN II and for their observed variation with the metallicity. In case no supernova occurs when a black hole is formed, single-star models can still account for more than half of the observed (SN Ib+SN Ic)/SN II ratio for $Z \ge$ Z$_\odot$. For the gamma ray burst rate, our models produce too large a number for such an event, even if we restrict the progenitor to the WO stars. This confirms that only a fraction of the WC / WO stars evolve toward gamma ray burst event, most likely those arising from stars that were initially very rapid rotators.}{}

\keywords {stars: Wolf-Rayet, supernovae: general, gamma rays: bursts}
\titlerunning{Progenitors of type Ib, Ic SNe and GRB}   

\maketitle

\section{Introduction}

Type Ib supernovae are core-collapse supernovae whose spectrum shows no hydrogen lines. The spectra of type Ic show no hydrogen and helium lines \citep[see e.g.][]{Wheeler1987a,Nomoto1994a}. The progenitors of these core collapse supernovae are thus believed to be stars stripped of their original H-rich envelope for type Ib's and also of their He-rich envelope for type Ic's. Progenitors are therefore Wolf-Rayet stars. The observed frequency at solar metallicity of type Ibc supernovae is about 20\% the frequency of type II supernovae \citep[see e.g.][]{Cappellaro1999a}, which represents a significant fraction of all core collapse supernovae.  In four cases, the typical spectrum of a type Ic supernova has been observed associated with a long soft gamma ray burst (GRB) event \citep{Woosley2006a}, indicating a privileged link between type Ic's and the most powerful supernova explosions observed in the Universe. That type Ibc represents a significant proportion of core collapse supernovae and the link between type Ic's and the long soft GRBs justify the study of the evolution leading to such events.

In this context, a very interesting feature observed by \citet{Prantzos2003a} is the increase with the metallicity in the number fraction of type Ibc to type II supernovae.  Such a trend is reproduced well by rotating single-star models accounting for metallicity-dependent stellar winds \citep{Meynet2003a}. On the other hand, a well known scenario explaining this trend comes from close binary-star evolution \citep[see for example][]{Podsiadlowski1992a,Vanbeveren2007a,Eldridge2008a}. In this scenario, the hydrogen-rich envelope is removed either through a Roche lobe overflow process or during a common envelope phase, producing a WR star.

In the present work, we take the opportunity of recent observations by \citet{Prieto2008a} from which separated frequencies for type Ib and Ic can be deduced to get a step further. The questions we want to address are: how many type Ib and Ic supernovae normalized to the number of type II supernovae can we expect from single-star models at different metallicities, how the numbers derived from single-star models compare with the observed values, and how the results change no supernova event occurs if every time a black hole forms. Answers to those questions will provide some hints as to the importance of the single star scenario in explaining the evolution leading to the type Ibc supernovae events and will provide useful quantities for more comparisons with similar outputs obtained from close binary evolution scenarios.

Other questions will be addressed, for instance, the absence of H and He lines does not necessarily imply that the associated elements are completely absent from the ejecta. It only tells us that the physical conditions (the chemical composition being only one of them) are such that no lines of these elements are formed at the time of the supernova event. Actually, the quantities of hydrogen and helium ejected by these two types of supernovae are not known. In this paper, guided by stellar models for single stars, we estimate the minimum quantities of hydrogen and of helium, we may expect in type Ib and type Ic supernovae. We also indicate the relations of filiation between type Ib, Ic supernovae and the WNL, WNE, WC, and WO stars.
 
The second section of this paper briefly recalls the main physical ingredients of the stellar models used in this study. Based on these models, we discuss in Sect.~\ref{Progen} the nature of the supernova progenitors, and in Sect.~\ref{Chemical} the chemical composition of the supernova ejecta. In Sect.~\ref{SN}, we discuss the link between the chemical composition of the ejecta and the nature of the supernova. The nature of the stellar remnant is discussed in Sect.~\ref{Rema}. Theoretical rates for different supernova types are estimated in Sect.~\ref{Freq} and compared with observations. Section \ref{GRB} discusses the implications for the long soft GRB progenitors and finally conclusions are presented in Sect.~\ref{Conc}.

\section{Stellar models}\label{Models}

All the rotating stellar models we use here come from \citet{Meynet2003a,Meynet2005a} (hereafter papers X and XI). A complete description of the physics used can be found in these two papers. We just recall here some important characteristics.

\begin{itemize}
\item All the models considered in this work are computed with an initial equatorial velocity of $300 \,\mathrm{km}\,\mathrm{s^{-1}}$. This initial velocity implies time-averaged equatorial velocities between 160 and 250 km s$^{-1}$ depending on the initial mass and metallicity (see tables 1 in papers X and XI). These values are in the range of observed values for OB stars; for instance, \citet{Huang2006a} present projected rotational velocities for 496 OB stars belonging to 19 young galactic clusters with estimated ages between 6 and 73 Myr. Mean $v\sin i$ values of 139, 154, and 151 km s$^{-1}$ were obtained for groups of O9.5-B1.5, B1.5-B5.0, and B5.0-B9.0 type stars. These authors derived the underlying probability distribution for the equatorial velocities $v$ and obtained a peak at 200 km s$^{-1}$. \citet{Dufton2006a} obtain a peak of $v$ at 250 km s$^{-1}$ with a full width half maximum of approximately 180 km s$^{-1}$ for the unevolved targets in the galactic clusters NGC 3293 and 4755.
\item Mass loss rates are those of \citet{Vink2000a,Vink2001a}, which take wind bistability into account. The rates of \citet{deJager1988a} are used outside the domain of application of the rates of \citet{Vink2000a,Vink2001a}. During the WR phase, the mass loss rates are those of \citet{Nugis2000a}. These mass loss rates, which account for the clumping effects in the winds, are lower by a factor of 2-3 than the ones used in our previous non-rotating stellar grids \citep{Meynet1994a}.
\item The dependence of mass loss on rotation \citep{Maeder2000a} was taken into account, as is the dependence on the metallicity. During the non--WR phases of the present models, we assumed that the mass loss rates depend on the initial metallicity as $\dot M(Z)=(Z/Z_\odot)^{1/2} \dot M(Z_\odot)$ \citep{Kudritzki2000a,Vink2001a}.
\end{itemize} 

Models were computed for four metallicities: $Z = 0.004$, $Z = 0.008$, $Z = 0.020$ (standard) and $Z = 0.040$. The initial masses considered are indicated in Table~\ref{HeMass}. All the models were computed up to the end of the core helium-burning phase. The very short durations of the advanced phase imply that the star at the end of the core He-burning phase has already reached its final mass. Moreover, the decoupling between the rapid evolution of the core governed by the neutrino emission and the slow evolution of the envelope does not allow the envelope to change a lot during the advanced phases. Thus we consider that the properties of the outer layers of our models obtained at the end of the core He-burning phase are very near the one that would have obtained had we pursued the computation until the presupernova stage\footnote{We checked using the solar metallicity models of \citet{Hirschi2004a}, which were pursued until the presupernova stage, that indeed this is the case.}.

In previous papers \citep[see e.g.][Paper X; Paper XI]{Meynet2000a,Maeder2001a,Maeder2008a}, we compared the outputs of the rotating models to many observed features of massive stars and showed that the models accounting for the effects of rotation gave much better agreement for fitting observed features such as
\begin{itemize}
\item surface enrichments,
\item the blue to red supergiant ratio in the SMC,
\item the variation with metallicity in the number ratio of WR to O-type stars,
\item the observed number ratio of WN to WC stars in the LMC and SMC,
\item the variations with metallicity in the number ratio of type Ibc to type II supernovae,
\item the existence of Wolf-Rayet stars showing both at their surface H- and He-burning products.
\end{itemize}

In papers X and XI we discussed the last four points above in details. Here, as indicated in the introduction, we focus on the nature of the progenitors for the type Ib and type Ic SNe, taken separately. 

\section{Nature of supernova progenitors}\label{Progen}

The nature of the supernova progenitors was determined by the surface composition of the star at the presupernova stage. The following classification scheme was adopted:

\begin{itemize}
\item All the stars not ending their lifetime as a Wolf-Rayet star end their nuclear life as a blue or a red supergiant. They are hereafter called supergiant (SG). 
\item The star is considered to be a WR star when $\log T_{\rm eff} > 4.0$ and the mass fraction of hydrogen at the surface $X_{\mathrm{S}}$ is inferior to 0.4 \citep[same criterion as in][]{Meynet2005a}.
\item Wolf-Rayet stars with hydrogen at their surface ($X_\mathrm{S} > 10^{-5}$) are WNL stars.
\item Wolf-Rayet stars without hydrogen on their surface ($X_\mathrm{S} < 10^{-5}$) and with surface carbon abundance inferior to nitrogen abundance are WNE stars.
\item Wolf-Rayet stars with carbon abundance on the surface superior to nitrogen abundance are WC or WO stars. To distinguish between WC and WO, we follow \citet{Smith1991a}: if the ratio of $\frac{\mathrm{C}+\mathrm{O}}{\mathrm{He}}$ (in number) is less than 1, we have a WC, otherwise a WO.
\end{itemize}

A star enters the WR phase as a WNL, then may evolve through the sequence WNL$\rightarrow$ WNE $\rightarrow$ WC $\rightarrow$ WO. A star can end its evolution at any of these stages. We do not consider the WN/WC phase here, which can occur as a very short transition phase between the WNE and WC phases.

\begin{table*}
\caption{For all our models, various quantities are given here (see text for details).}
\label{HeMass}
\begin{center}
\begin{tabular}{ c c c c c c c c || c c c c c c c c}
\hline\hline
$M_\mathrm{ini} $ & $M_\mathrm{end\,He} $ & $M_\mathrm{He} $ & $M_\mathrm{CO} $ & $M_\mathrm{rem} $ & Prog. & SN & Rem. & $M_\mathrm{ini}$ & $M_\mathrm{end\,He}$ & $M_\mathrm{He}$ & $M_\mathrm{CO}$ & $M_\mathrm{rem}$ & Prog. & SN & Rem.\\
\hline
& & & & & & & & & & & & & & &\\
\multicolumn{8}{c ||}{$\mathbf{Z = 0.004}$} & \multicolumn{8}{c}{$\mathbf{Z = 0.020}$}\\
$12$ & $11.8$ & $3.74$ & $1.78$ & $1.4$ & SG & II & NS & $12$ & $10.5$ & $3.81$ & $1.94$ & $1.4$ & SG & II & NS\\
$15$ & $14.1$ & $5.01$ & $2.84$ & $1.7$ & SG & II & NS & $15$ & $10.2$ & $5.97$ & $3.61$ & $1.7$ & SG & II & NS\\
$20$ & $18.0$ & $7.45$ & $4.79$ & $2.0$ & SG & II & NS & $20$ & $11.8$ & $9.01$ & $6.62$ & $2.0$ & SG & II & NS\\
$25$ & $20.0$ & $9.95$ & $7.06$ & $2.5$ & SG & II & NS & $25$ & $11.3$ & $11.33$ & $8.98$ & $2.5$ & WNL & II & NS\\
$30$ & $18.9$ & $14.16$ & $11.14$ & $2.9$ & SG & II & BH & $40$ & $12.7$ & $12.70$ & $11.75$ & $3.0$ &  WC & Ic & BH\\
$40$ & $22.3$ & $21.59$ & $17.16$ & $3.5$ & WNL & II & BH & $60$ & $14.6$ & $14.60$ & $13.67$ & $3.6$ & WC & Ic & BH\\
$60$ & $28.5$ & $28.50$ & $28.46$ & $4.4$ & WO & Ic & BH & $85$ & $12.3$ & $12.30$ & $11.22$ & $3.0$ & WC & Ic & BH\\
$120$ & $17.2$ & $17.20$ & $17.18$ & $3.4$ & WC & Ic & BH & $120$ & $11.3$ & $11.30$ & $10.40$ & $2.8$ & WC & Ic  & BH\\
\hline
& & & & & & & & & & & & & & & \\
\multicolumn{8}{c ||}{$\mathbf{Z = 0.008}$} & \multicolumn{8}{c}{$\mathbf{Z = 0.040}$}\\
$20^1$ & $15.5$ & $-$ & $4.85$ & $2.1$ & SG & II & NS & $20$ & $9.2$ & $9.05$ & $6.84$ & $1.4$ & SG & II & NS\\
$25^1$ & $14.0$ & $-$ & $6.92$ & $2.7$ & SG & II & NS & $25$ & $9.6$ & $9.60$ & $8.19$ & $1.7$ & WC & Ib & NS\\
$30$ & $12.1$ & $12.10$ & $11.10$ & $2.9$ & WC & Ic & BH & $40$ & $11.4$ & $11.40$ & $10.48$ & $2.8$ & WC & Ic & BH\\
$40$ & $17.4$ & $17.40$ & $17.23$ & $3.5$ & WC & Ic & BH & $85$ & $7.2$ & $7.20$ & $6.07$ & $2.2$ & WC & Ic & NS\\
$60$ & $16.4$ & $16.40$ & $16.44$ & $3.3$ & WO & Ic & BH & $120$ & $7.1$ & $7.10$ & $6.10$ & $2.2$ & WC & Ic & NS\\
$120$ & $13.4$ & $13.40$ & $13.24$ & $3.0$ & WC & Ic & BH & & & & & & &\\
\hline
& & & & & & & & & & & & & & & \\
\multicolumn{8}{c ||}{\textbf{Homogenous model at} $\mathbf{Z = 0.002}$} & & & & & & & &\\
$60^2$ & $27.8$ & $-$ & $-$ & $4.0$ & WC & Ib &BH & & & & & & &\\
\cline{1-8}
\multicolumn{16}{l}{(1) Non-rotating models from \citet{Meynet1994a} used to estimate the minimum initial mass of stars ending their lifetime}\\
\multicolumn{16}{l}{in the WNL phase at that metallicity. These models were computed with artificially enhanced mass loss rates, which}\\
\multicolumn{16}{l}{somewhat compensates for the neglect of the effects of rotation.}\\

\multicolumn{8}{l}{(2) Model computed up to the end of silicon burning.} & & & & & & &\\
\end{tabular}
\end{center}
Note: All masses are given in $[\mathrm{M}_\odot]$. The SN type is given for $m^{\rm min}_{\rm H}=0$, $m^{\rm min}_{\rm He}=0.6$ and supposing that a SN occurs even when a BH is formed (see text). The remnant masses used are those given by \citet{Hirschi2005a}, except for the homogenous models (see text).
\end{table*}

Table \ref{HeMass} gives for the 4 metallicities considered here the initial mass of the star $M_\mathrm{ini}$, the mass at the end of the helium-burning phase $M_\mathrm{end\, He}$, the mass of the He-core, $M_\mathrm{He} $, defined as the mass interior to which the mass fraction of helium (or of its products after fusion) is higher than 0.75, the mass of the carbon-oxygen core, $M_\mathrm{CO} $, defined as the mass interior to which the mass fraction of carbon plus oxygen is higher than 0.75, the mass of the remnant $M_\mathrm{rem}$ \citep[given by][]{Hirschi2005a}, the type of the progenitor, the SN type produced (see Sect.~\ref{SN}), and the remnant type (see Sect.~\ref{Rema}).

\begin{figure}[t]
\centering
\includegraphics[angle=0,width=9cm]{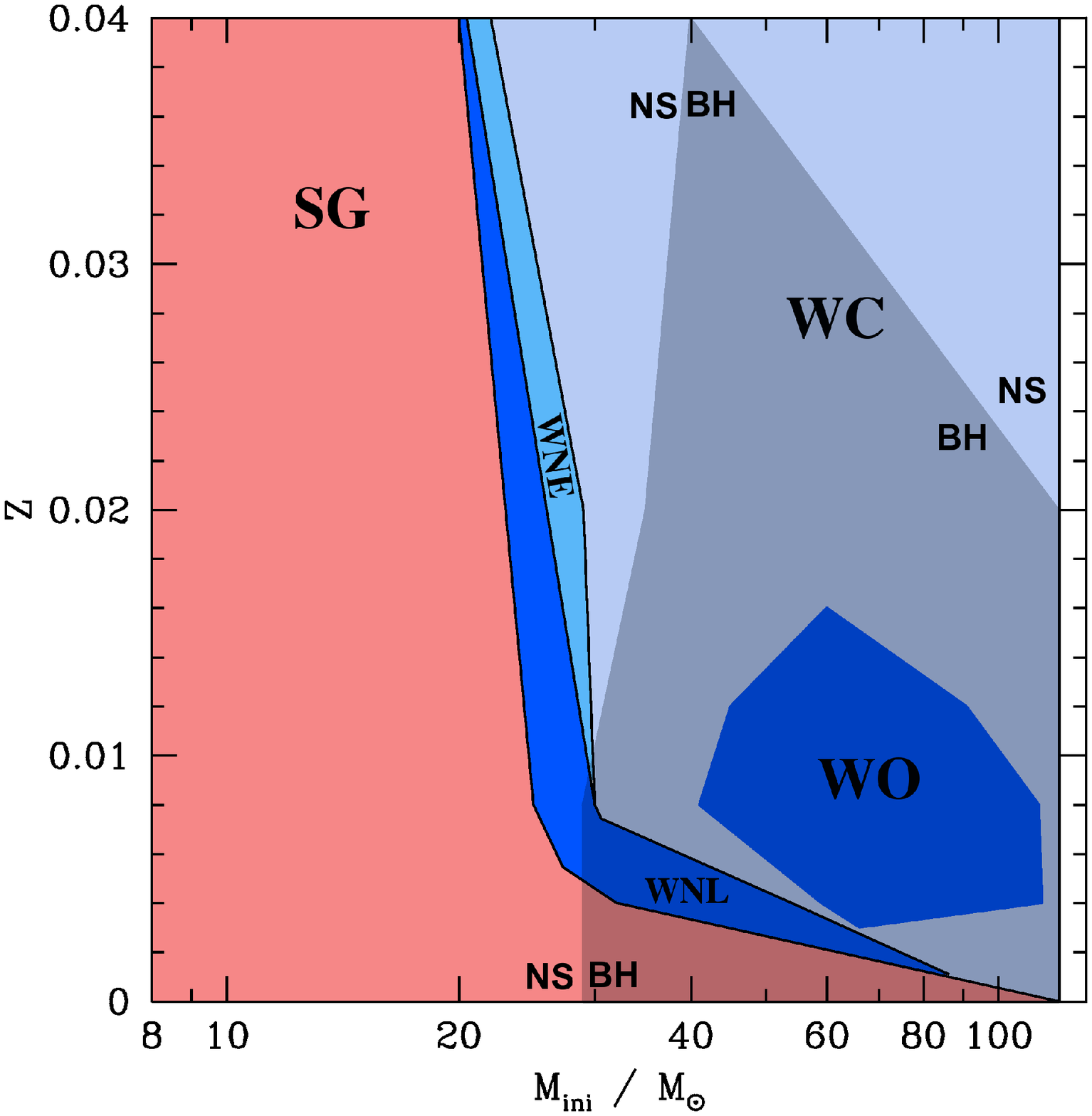}
\caption{Ranges of masses of different types of supernova progenitors at different metallicities. For each area, the type of progenitor is indicated in the figure (see text). The shading on the right indicates the area where formation of a black hole (BH) is expected ; elsewhere, the remnant is a neutron star (NS) (color figure available online).}
\label{WRtype}
\end{figure}

In Fig.~\ref{WRtype}, we show how the initial mass ranges of various supernova progenitors vary as a function of the metallicity. To build this figure we had to adopt a method to determine the mass limits between the different zones. To illustrate, let us consider one specific case. At $Z=0.02$, from Table~\ref{HeMass} we know that the 25 M$_\odot$ stellar model ends its lifetime as a WNL star and the 40 M$_\odot$ as a WC star, but we know neither at which mass the transition from WNL to WC star occurs nor whether some stars will end their life in the WNE stage. To find the transition between the WNL and WNE stage, we deduce from the chemical structure of our 25 M$_\odot$ at the end of its evolution, the mass that should be removed in order for the star to become a WNE star (in that case 0.35 M$_\odot$ or 1.4\% of its total initial mass). From our 40 M$_\odot$ stellar model at the end of its evolution, we deduce the mass it has lost since its entry into the WNE phase (1.53 M$_\odot$ or 6.1\% of its initial mass). Assigning, by convention, a  minus sign to this last quantity (mass lost) and a positive sign to the former one (mass which has to be lost), plotting these two quantities in a plane  ``mass-to be/or- lost'' versus ``initial mass'', the abscissa of the intersection of the line connecting these two points with the horizontal line at ordinate 0 gives the initial mass  for which the endpoint of the evolution just corresponds to the stage where ``no mass'' has to be lost to enter into the WNE phase and where no mass has been lost since the entry into that phase. Thus this is the minimum initial mass to end its evolution as a WNE star. Using similar procedures for all the limits, we obtain the regions shown in Fig.~\ref{WRtype}. Note also that we used the results of \citet{Ekstrom2008a} for complementing the figure with Pop III stellar models.

Looking at that figure, we can make the following remarks.
\begin{itemize}
\item As expected, the lower mass limit for having a WR progenitor decreases when the metallicity increases. We see that the dependence on the metallicity of this limit is much stronger at low $Z$ than at high $Z$.
\item The mass range of stars ending their life as a WN star is relatively narrow. Below this mass range, the stars do not succeed in entering the WR phase; above it, mass loss rates are efficient enough for allowing the star to evolve further into the WC or WO phase.
\item We note that at metallicities higher than about 0.02, the WN progenitors are more or less equally distributed among WNL and WNE stars, while for metallicities below 0.01 all WN progenitors are WNL stars. Actually present stellar models predicts such a narrow range of masses for WNE progenitors below $Z=0.008$ that we neglected it. This can be explained in the following way: first let us point out that WNE stars can be produced only when the star has evolved beyond the core H-burning phase. Indeed, regions without hydrogen are present in the star only after the main-sequence phase has ended. At high metallicity, because of strong stellar winds, the outer part of the H-burning core is uncovered at an early stage when the size of the He-burning core is still small and has not yet transformed a significant part of the former H-burning core into carbon and oxygen. This allows the intermediate region between the He-burning core and the H-burning shell to have a large extension in mass. Thus, the WNE phase has a significant duration \citep[see Fig.~9 of][]{Meynet2005a}, which allows a narrow range of initial masses for which the endpoint of the evolution occurs when the star is in the WNE phase. At low metallicity, the outer part of the H-burning core is uncovered at later evolutionary stages when the He-burning core has grown in mass, reducing the extent in mass of the region with chemical composition typical of the WNE phase. This reduces the duration of the WNE phase and also the initial mass range of stars ending their evolution in that stage.
\item Above 30 M$_\odot$ and for metallicities higher than 0.008, all stars end their life as WC/WO stars. For metallicities below 0.008, the minimum initial mass for stars ending their lifetime as WC stars rapidly increases when the metallicity becomes lower. 
\item The WO progenitors region appears as an ``island'' in the ``WC'' sea, located at low metallicity. As first explained by \citet{Smith1991a}, the low metallicity position of this island comes from the fact that, in order to form a WO star, the He-burning core must be uncovered at a very advanced stage of its evolution when a lot of helium has been converted into carbon and oxygen. This is more easily realized at low metallicity where the mass loss rates are lower, hence the core is revealed at a later stage. We also note that the WO progenitors do arise neither from the most massive nor from the less massive stars having their He-burning core uncovered. This reflects the delicate interplay between two counteracting effects. On one hand, in order for a star to become a WO star, strong enough stellar winds are necessary; on the other hand, as explained above, the stellar winds should not be too strong, otherwise no WO stars but instead WC stars are formed!
\end{itemize}

\begin{figure}[t]
\centering
\includegraphics[angle = 0, width=9cm]{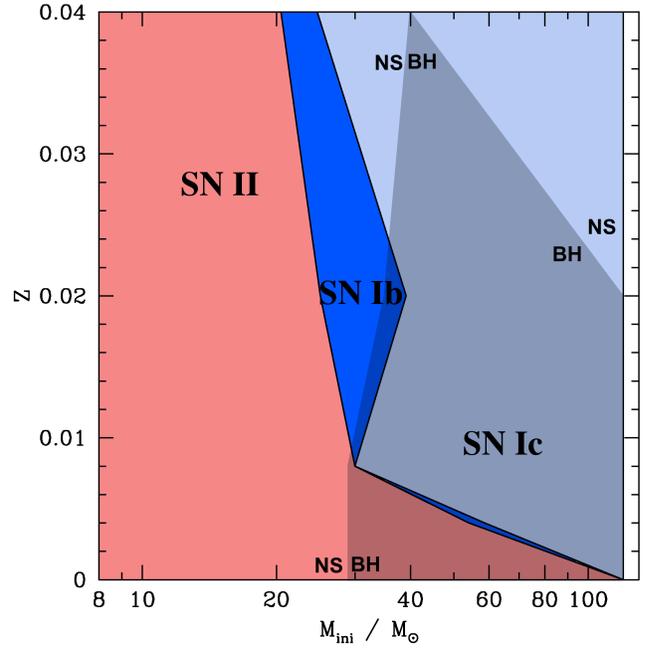}
\caption{Same as Fig.~\ref{WRtype} for various types of SNe. The SN type is given for $m^{\rm min}_{\rm H}=0$, $m^{\rm min}_{\rm He}=0.6$ and supposing that an SN occurs even when a BH is formed (see text).}
\label{SNtype}
\end{figure}

\begin{table*}
\caption{Initial mass $M_\mathrm{ini}$, total released mass $M_\mathrm{rel}$, mass and fraction of ejected H ($M_\mathrm{H}$), He ($M_\mathrm{He}$), C ($M_\mathrm{C}$), N ($M_\mathrm{N}$), O ($M_\mathrm{O}$) and of heavy elements ($M_\mathrm{heavy}$), with all masses in $\mathrm{M}_\odot$.}
\label{EjMass}
\centering
\begin{tabular}{ c c c c c c c c}
\hline\hline
$M_\mathrm{ini} $ &  $m_\mathrm{rel}$ & $m_\mathrm{H}$ & $m_\mathrm{He}$ & $m_\mathrm{C}$ & $m_\mathrm{N}$ & $m_\mathrm{O}$ & $m_\mathrm{heavy}$\\
\hline
& & & & & & &\\
\multicolumn{8}{c}{$\mathbf{Z = 0.004}$}\\
$12$ & $10.4$ & $5.52\,(53\%)$ & $4.30\,(41\%)$ & $0.15\,(1.4\%)$ & $1.7\cdot 10^{-2}\,(0.16\%)$  & $8.9\cdot 10^{-2}\,(0.86\%)$ & $0.58\,(5.6\%)$\\
$15$ & $12.4$ & $6.11\,(49\%)$ & $4.94\,(40\%)$ & $0.25\,(2.0\%)$ & $1.5\cdot 10^{-2}\,(0.12\%)$ & $0.42\,(3.4\%)$ & $1.35\,(11\%)$\\
$20$ & $16.0$ & $6.57\,(41\%)$ & $6.35\,(40\%)$ & $0.49\,(3.1\%)$ & $1.8\cdot 10^{-2}\,(0.11\%)$  & $1.41\,(8.8\%)$& $3.08\,(19\%)$\\
$25$ & $17.5$ & $5.48\,(31\%)$ & $7.09\,(41\%)$ & $0.88\,(5.0\%)$ & $2.1\cdot 10^{-2}\,(0.12\%)$ & $2.64\,(15\%)$ & 4.93\,(28\%)\\
$30$ & $16.0$ & $1.58\,(9.9\%)$ & $5.75\,(36\%)$ & $1.45\,(9.0\%)$ & $2.1\cdot 10^{-2}\,(0.13\%)$  & $5.07\,(32\%)$ & $8.67\,(54\%)$\\
$40$ & $18.8$ & $0.45\,(2.4\%)$ & $4.35\,(23\%)$ & $2.11\,(11\%)$ & $1.4\cdot 10^{-2}\,(0.07\%)$ & $9.93\,(53\%)$ & $14.0\,(75\%)$\\
$60$ & $24.1$ & $0.00\,(0.0\%)$ & $0.41\,(1.7\%)$ & $2.16\,(9.0\%)$ & $1.4\cdot 10^{-6}\,(0.0\%)$  & $14.44\,(60\%)$& $23.69\,(98\%)$\\
$120$ & $13.8$ & $0.00\,(0.0\%)$ & $0.48\,(3.5\%)$ & $1.73\,(13\%)$ & $0.00\,(0.0\%)$  & $6.92\,(50\%)$& $13.32\,(97\%)$\\
\\
\multicolumn{8}{c}{$\mathbf{Z = 0.008}$}\\
$20^1$ & $13.4$ & $9.70\,(73\%)$ & $3.55\,(27\%)$ & $2.4\cdot 10^{-2}\,(0.18\%)$ & $7.5\cdot 10^{-3}\,(0.06\%)$ & $5.6\cdot 10^{-2}\,(0.42\%)$ & $0.11\,(0.8\%)$\\ 
$25^1$ & $11.3$ & $7.83\,(69\%)$ & $3.40\,(30\%)$ & $1.6\cdot 10^{-2}\,(0.14\%)$ & $1.5\cdot 10^{-2}\,(0.14\%)$& $4.3\cdot 10^{-2}\,(0.38\%)$ & $0.91\,(0.8\%)$\\
$30$ & $9.2$ & $0.00\,(0.0\%)$ & $0.54\,(5.9\%)$ & $1.06\,(12\%)$ & $1.8\cdot 10^{-3}\,(0.02\%)$ & $3.30\,(36\%)$ & $8.66\,(94\%)$\\
$40$ & $13.9$ & $0.00\,(0.0\%)$ & $0.31\,(2.2\%)$ & $1.75\,(13\%)$ & $3.0\cdot 10^{-4}\,(0.0\%)$  & $6.98\,(50\%)$& $13.59\,(98\%)$\\
$60$ & $13.1$ & $0.00\,(0.0\%)$ & $0.49\,(3.7\%)$ & $1.68\,(13\%)$ & $0.00\,(0.0\%)$ & $6.48\,(49\%)$ & $12.61\,(96\%)$\\
$120$ & $10.4$ & $0.00\,(0.0\%)$ & $0.44\,(4.2\%)$ & $1.36\,(13\%)$ & $0.00\,(0.0\%)$ & $4.55\,(44\%)$ & $9.96\,(96\%)$\\
\\
\multicolumn{8}{c}{$\mathbf{Z = 0.020}$}\\
$12$ & $9.1$ & $4.13\,(45\%)$ & $4.20\,(46\%)$ & $0.18\,(2.0\%)$ & $5.4\cdot 10^{-2}\,(0.59\%)$ & $0.18\,(1.9\%)$ & $0.77\,(8.5\%)$\\
$15$ & $8.5$ & $2.53\,(30\%)$ & $3.83\,(45\%)$ & $0.32\,(3.7\%)$ & $5.0\cdot 10^{-2}\,(0.59\%)$ & $0.84\,(9.9\%)$ & $2.14\,(25\%)$\\
$20$ & $9.8$ & $1.40\,(14\%)$ & $3.51\,(36\%)$ & $0.74\,(7.6\%)$ & $4.7\cdot 10^{-2}\,(0.48\%)$ & $2.22\,(23\%)$ & $4.89\,(50\%)$\\
$25$ & $8.8$ & $3.78\cdot 10^{-5}\,(0.0\%)$ & $2.04\,(23\%)$ & $1.13\,(13\%)$ & $2.6\cdot 10^{-2}\,(0.29\%)$ & $3.50\,(40\%)$ & $6.76\,(77\%)$\\
$40$ & $9.7$ & $0.00\,(0.0\%)$ & $0.53\,(5.5\%)$ & $1.21\,(12\%)$ &  $3.8\cdot 10^{-6}\,(0.0\%)$ & $3.77\,(39\%)$ & $9.17\,(95\%)$\\
$60$ & $11.0$ & $0.00\,(0.0\%)$ & $0.54\,(4.9\%)$ & $1.43\,(13\%)$ & $4.4\cdot 10^{-6}\,(0.0\%)$  & $4.86\,(44\%)$& $10.46\,(95\%)$\\
$85$ & $9.3$ & $0.00\,(0.0\%)$ & $0.53\,(5.7\%)$ & $1.11\,(12\%)$ & $3.7\cdot 10^{-6}\,(0.0\%)$  & $3.43\,(37\%)$& $8.77\,(94\%)$\\
$120$ & $8.5$ & $0.00\,(0.0\%)$ & $0.49\,(5.8\%)$ & $0.96\,(11\%)$ &  $3.4\cdot 10^{-6}\,(0.0\%)$ & $2.94\,(35\%)$& $8.01\,(94\%)$\\
\\
\multicolumn{8}{c}{$\mathbf{Z = 0.040}$}\\
$20$ & $7.8$ & $7.1\cdot 10^{-2}\,(0.91\%)$ & $1.96\,(25\%)$ & $0.77\,(9.9\%)$ & $4.62\cdot 10^{-2}\,(0.59\%)$ & $2.29\,(29\%)$ & $5.77\,(74\%)$\\
$25$ & $7.9$ & $0.00\,(0.0\%)$ & $0.51\,(6.5\%)$ & $0.65\,(8.2\%)$ & $0.00\,(0.0\%)$ & $1.88\,(24\%)$ & $7.39\,(94\%)$\\
$40$ & $8.6$ & $0.00\,(0.0\%)$ & $0.41\,(8.3\%)$ & $1.00\,(12\%)$ & $0.00\,(0.0\%)$ & $3.07\,(36\%)$ & $7.89\,(92\%)$\\
$85$ & $5.0$ & $0.00\,(0.0\%)$ & $0.48\,(9.6\%)$ & $0.33\,(6.6\%)$ & $0.00\,(0.0\%)$ & $0.89\,(18\%)$ & $4.52\,(90\%)$\\
$120$ & $4.9$ & $0.00\,(0.0\%)$ & $0.44\,(9.0\%)$ & $0.33\,(6.8\%)$ & $0.00\,(0.0\%)$ & $0.90\,(18\%)$ & $4.46\,(91\%)$\\
\hline\\
\multicolumn{8}{c}{\textbf{Homogenous model at} $\mathbf{Z = 0.002}$}\\
$60^2$ & $23.8$ & $0.00\,(0.0\%)$ & $0.64\,(2.7\%)$ & $2.33\,(9.8\%)$ & $3.66\cdot 10^{-7}\,(0.0\%)$ & $15.56\,(65\%)$ & $23.12\,(97\%)$\\
\hline
\multicolumn{8}{l}{(1) See note (1) in Table~\ref{HeMass}.}\\
\multicolumn{8}{l}{(2) Model computed up to the end of silicon burning.}\\
\end{tabular}
\end{table*}

\begin{table*}
\caption{Mass limits between type II and type Ib SNe ($M_\mathrm{lim}^\mathrm{II/Ib}$), and between type Ib SNe and Ic SNe ($M_\mathrm{lim}^\mathrm{Ib/Ic}$) in $\mathrm{M}_\odot$ (see text).}
\label{LimitMass}
\centering
\begin{tabular}{c c c c||c c c}
\hline\hline
& \multicolumn{3}{c||}{mass limit between type II -- type Ib SNe} & \multicolumn{3}{c}{mass limit between type Ib -- type Ic SNe}\\
$Z$ & $m^\mathrm{min}_\mathrm{H} = 0.00$ & $m^\mathrm{min}_\mathrm{H} = 0.25$ & $m^\mathrm{min}_\mathrm{H} = 0.50$ & $m^\mathrm{min}_\mathrm{He} = 0.4$ &  $m^\mathrm{min}_\mathrm{He} = 0.5$ &  $m^\mathrm{min}_\mathrm{He} = 0.6$\\
\hline
$0.004$ & $54.0$ & $48.1$ & $39.4$ & -- & $59.4$ & $58.6$ \\
$0.008$ & $30.0$ & $29.9$ & $29.7$ & $35.7 - 49.1$ & $31.6$ & $30.0$\\
$0.020$ & $25.1$ & $24.2$ & $23.1$ & -- & $110.2$ & $39.1$\\
$0.040$ & $20.5$ & $20.5$ & $20.5$ & -- & $26.2$ & $24.7$\\
\hline
\end{tabular}
\end{table*}

\begin{table*}
\caption{Mass ranges for various supernova progenitors, supernova types, and compact remnants at different metallicities (in $\mathrm{M}_\odot$).}
\label{masslimits}
\centering
\begin{tabular}{c c c c c c c c c c c}
\hline\hline 
Z & SG & WNL & WNE & WC & WO & Type II & Type Ib & Type Ic & NS & BH  \\
\hline\\
\multicolumn{11}{c}{\textbf{Reference case}}\\
$0.004$ & $8 - 32$ & $32 - 54$ & $-$ & $54 - 59;114 - 120$ & $59 - 114$ & $8 - 54$ & $54 - 59$ & $59 - 120$ & $8 - 29$ & $29 - 120$\\
$0.008$ & $8 - 25$ & $25 - 30 $ & $ - $ &$30 - 41; 113 - 120$ & $41 - 113$ & $8 - 30$ &  -  & $30 - 120$ & $8 - 29$ & $29 - 120$\\
$0.020$ & $8 - 23$ & $23 - 26$ & $26 - 29$ & $29 - 120$ & $-$ & $8 - 25$ & $25 - 39$ & $39 - 120$ & $8 - 35$ & $35 - 120$\\
$0.040$ & $8 - 20$ & $20 - 20.5$ & $20.5 - 22$ & $22 - 120$ & $-$ & $8 - 20.5$ & $20.5 - 25$ & $25 - 120$ & $8 - 120$ & --\textsuperscript{1}\\
\hline\multicolumn{11}{l}{(1) At this metallicity, only the $40\, \mathrm{M}_\odot$ model produces a BH.}
\end{tabular}
Note: The SN type is given for $m^{\rm min}_{\rm H}=0$, $m^{\rm min}_{\rm He}=0.6$ and supposing that an SN occurs even when a BH is formed (see text for more details).
\end{table*}

\section{Chemical composition of the ejecta}\label{Chemical}

The masses of hydrogen and helium ejected at the time of the explosive event are the key quantities for determining the type of the supernova event. These quantities are given for each of our stellar  models in Table~\ref{EjMass},which indicates for each metallicity and initial mass considered the total mass ejected at the time of the supernova explosion (column 2), the mass of hydrogen, of helium, of carbon, nitrogen and oxygen, and of heavy elements (columns 3 to 8).  Heavy elements are the sum of the mass ejected under the form of elements heavier than hydrogen and helium. For carbon and oxygen, we used relations given by \citet{Maeder1992a} between the mass of the CO core and the ejected mass of these two elements at the time of the supernova event.
 
\begin{figure}
\centering
\includegraphics[angle=0,width=9cm]{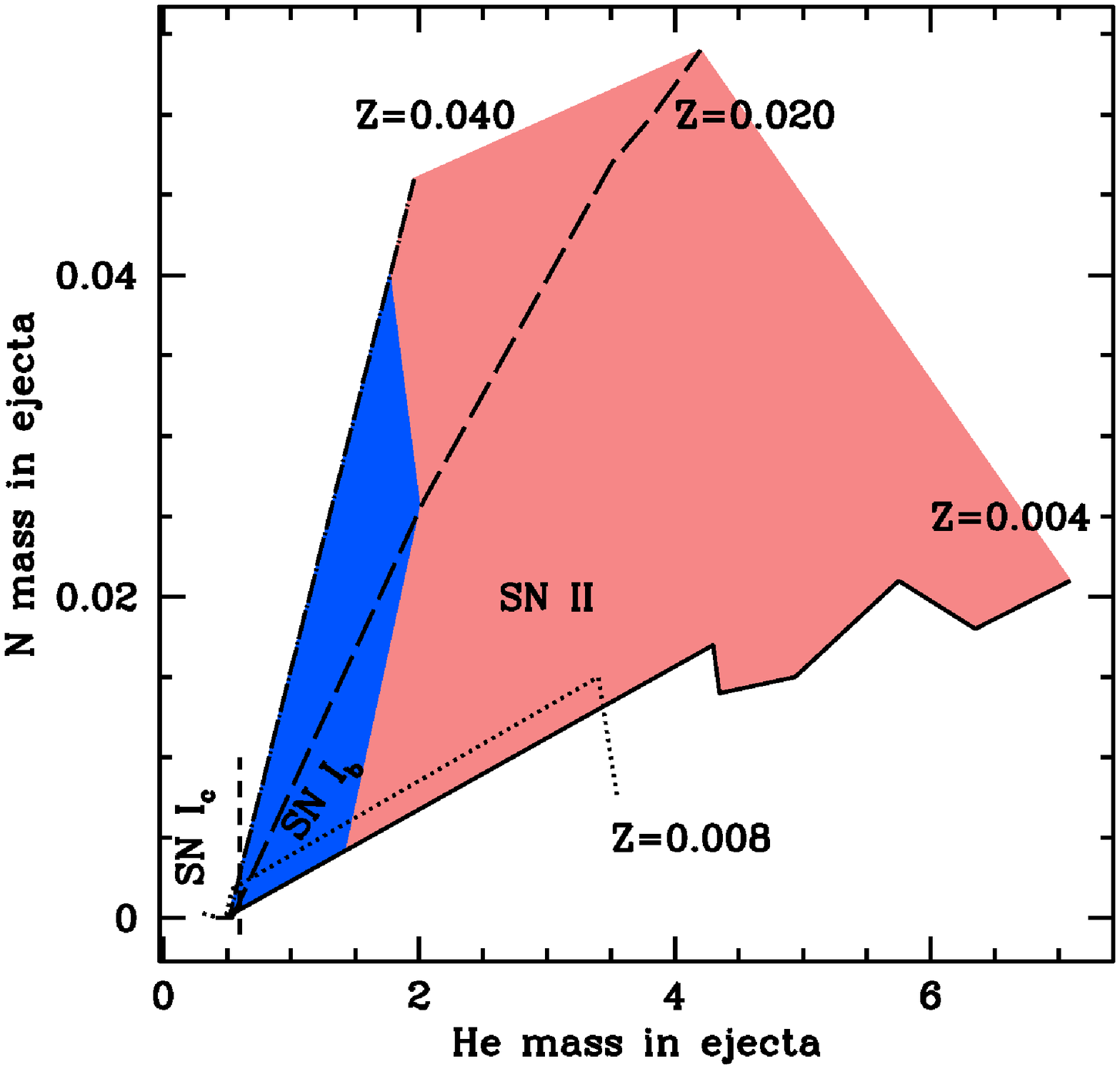}
\caption{Correlation between the mass of nitrogen in the supernova ejecta and the mass of helium.
The SN II region (red) and SN Ib region (blue) are represented, as well as the SN Ic region (on the left of the short dashed line). Masses are given in $\mathrm{M}_\odot$ (color figure available online).}
\label{HEN}
\end{figure}

From the results shown in Table~\ref{EjMass}, we can deduce the following trends.
\begin{itemize}
\item Most of supergiant progenitors eject more than 70\% of their ejecta in the form of hydrogen and helium. At any given metallicity, the proportion of the ejecta under the form of hydrogen and helium is greater in lower initial mass stars.
\item The 20 M$_\odot$ stellar model at $Z=0.04$ is a special case of supergiant progenitor in the sense that it only ejects small amounts of hydrogen and helium.  This model is a blue supergiant ($\log T_{\rm eff}=4.338$) at the end of its stellar lifetime and is actually on the verge of becoming a Wolf-Rayet star. It should lose only 0.09 M$_\odot$ to uncover layers with a hydrogen mass fraction inferior to 0.4. Probably a slightly more massive star would show an evolutionary path of the type O--red SG--blue SG--WR. Let us mention here that the single-star evolutionary scenario predicts a strong anti-correlation in the number of RSG and of WR stars in single-aged populations \citep{Leitherer1999a}. Close binary evolution may produce WR stars more easily from lower initial mass stars \citep{Podsiadlowski1992a} and thus explain the simultaneous presence of RSG and WR stars even in stellar populations with age spread inferior to a few Myr.
It would be interesting to obtain from well-observed, single-aged stellar populations quantitative constraints on the occurrence of RSG and WR stars and to compare them with population synthesis models accounting for single and close binary evolution.
\item Our computed WNL progenitors eject less than 1 M$_\odot$ of hydrogen, but more than 1 M$_\odot$ of helium (more generally every time some hydrogen is present in the ejecta, the mass of ejected helium is above $\sim$2 M$_\odot$). They eject greater quantities of carbon and oxygen than supergiant progenitors. About 3/4 of the mass of the ejecta is in the form of  heavy elements.
\item WC and WO progenitors eject more than 90\% of their ejecta in the form of heavy elements. In these cases, the mass of ejected helium is between 0.31 and 0.54 M$_\odot$. 
\item We see that the minimum helium content in the ejecta of a core collapse supernova is about $0.3\,\mathrm{M}_\odot$.  This means in particular that type Ic SNe should at least contain this minimum amount of helium in their ejecta. This agrees with the results of non rotating models presented in \citet{Eldridge2004a}, and with binary models of Eldridge (private communication). There is to date no clear measurement of the mass of helium contained in the ejecta of type Ic SNe \citep[although see][]{Elmhamdi2006a}. It would be very interesting to find a correlation or an anticorrelation between the mass of ejected helium and the mass of another element produced in the same region of the star that would be  more easily detectable. Typically, such correlations exist between the ejected mass of He and of N, as can be seen in Fig.~\ref{HEN}. We see that the correlation depends on the metallicity so that to use such theoretical guideline for the He-content, some measurement of the metallicity is needed (such correlations would allow an idea of the metallicity in case the amounts of helium and of nitrogen in the ejecta can be measured!). To advance on the question of the He-content in type Ic supernova ejecta, we would need similar correlations but of course with another element, since in type Ic's the amount of nitrogen is zero. Probably $^{22}$Ne would be an interesting candidate, since it traces the He-rich regions of the star. The difficulty would then be to distinguish the $^{22}$Ne abundances from that of $^{20}$Ne in the spectrum analysis. In  a future work, we shall study this question.
\item There is no difference in the chemical composition of the ejecta of the WC and the WO stars.
\end{itemize}

\section{The supernova types}\label{SN}

As already emphasized in the introduction, the conditions for producing type Ib and type Ic supernovae are based on the presence/absence of H, He lines in the supernova spectrum. Here we adopt a criterion based on the quantities of hydrogen and of helium ejected at the time of the supernova event to link a given chemical structure of the supernova progenitor to a given supernova type. Let us call $m^\mathrm{min}_\mathrm{H}$ the minimum quantity of hydrogen that should be present in the ejecta for the spectrum of the supernova to present H lines and $m^\mathrm{min}_\mathrm{He}$, the minimum quantity of helium that should be present in the ejecta for the spectrum of the supernova to present He lines. In that case a given initial mass star $M$ presenting at the end of its evolution a quantity $m_\mathrm{H}(M)$ of hydrogen in its envelope, respectively $m_\mathrm{He}(M)$, of helium will  produce 
\begin{itemize}
\item a type II supernova if $m_\mathrm{H}(M) >  m^\mathrm{min}_\mathrm{H}$;
\item a type Ib supernova if $m_\mathrm{H}(M) <  m^\mathrm{min}_\mathrm{H}$ and $m_\mathrm{He}(M) >  m^\mathrm{min}_\mathrm{He}$;
\item a type Ic supernova if $m_\mathrm{H}(M) <  m^\mathrm{min}_\mathrm{H}$ and $m_\mathrm{He}(M) <  m^\mathrm{min}_\mathrm{He}$.
\end{itemize}
The quantities $m^\mathrm{min}_\mathrm{H}$ and $m^\mathrm{min}_\mathrm{He}$ are not known a priori. In the case of $m^\mathrm{min}_\mathrm{H} = 0$, we obtain the limiting masses, $M_{\rm lim}^\mathrm{II/Ib}$, indicated in the second column of Table 4, where stars with initial masses below $M_{\rm lim}^\mathrm{II/Ib}$ explode as type II supernovae and stars with masses above $M_{\rm lim}^\mathrm{II/Ib}$ explode as type Ib or type Ic supernovae. We see that, when the metallicity becomes higher, this mass limit decreases as expected. Interestingly also, we see that adopting $m^\mathrm{min}_\mathrm{H}=0.25$, or 0.5 M$_\odot$, pushes M$_{\rm lim}^\mathrm{II/Ib}$ towards lower values, thereby decreasing the mass range for type II supernovae and increasing it for type Ibc supernovae. Therefore $m^\mathrm{min}_\mathrm{H}=0$ leads to a lower limit for the number fraction of type Ibc to type II supernovae. Except at the metallicity $Z=0.004$, the changes remain quite modest. In the following we adopt the value $m^\mathrm{min}_\mathrm{H}=0$ as our reference value keeping in  mind that it may underestimate the number fraction of type Ibc to type II supernovae.

The choice of $m^\mathrm{min}_\mathrm{He}$ is more delicate. We first note that there are no models without any helium in the ejecta. For all the metallicities and masses considered here, the minimum value of the mass of helium in the ejecta is 0.31 M$_\odot$. Second, we see that passing from the value 0.4 to 0.5 changes the mass limit a lot between the type Ib and the type Ic. Still, some changes are brought when one passes from 0.5 to 0.6, but in general are much more modest except at solar metallicity. Choosing values for $m^\mathrm{min}_\mathrm{He}$ between 0.6 and 1 - 1.5 M$_\odot$ would not change the results significantly when adopting 0.6. From these considerations it appears that 0.6 is a kind of limiting value, because below that value, the mass limit between type Ib and type Ic presents an high sensitivity to the exact value adopted for $m^\mathrm{min}_\mathrm{He}$ (for instance, choosing 0.4 would make type Ic appearing only in a narrow interval of metallicities). Above it the sensitivity of the mass is much weaker and the mass limits are much less dependent on the exact value chosen. In that respect the value of 0.6 does appear to us as the most reasonable choice. Also this choice  gives a monotonic decrease in the mass limit between type Ib and type Ic when the metallicity increases and allows all WC and WO stars to be type Ic progenitors. Thus, in the following we adopt the limiting values corresponding to $m^\mathrm{min}_\mathrm{He}=0.6$. 

We are now able to determine the type of the supernova it will give for each model (see column 7 of Table~\ref{HeMass}). Table ~\ref{masslimits} gives for each metallicity the initial mass ranges leading to different progenitors, SN events, and stellar remnants. A comparison between the mass ranges for the different types of  progenitors and of SN types shows that all SG stars produce a type II SN. The WNL stars generally lead to type II SNe except may be those WNL stars on the verge of becoming WNE stars. They present nearly no hydrogen on their surface (see the case of 25 M$_\odot$ model at $Z=0.02$). WNE stars contain no hydrogen, but enough helium to become a type Ib SN. WC stars end their life as type Ic except the less massive ones that contain too much helium to be classified as a type Ic SN and thus would explode as a type Ib. Finally, all the WO stars lead to a type Ic SN event.

Using Table ~\ref{masslimits}, we can build Fig. ~\ref{SNtype}, showing the different supernova types expected for various initial masses and metallicities. As for Fig.~\ref{WRtype}, we use the models of \citet{Ekstrom2008a} to complete the figure for $Z = 0.0$. We see that type Ib SNe arise from a mass band recovering the WNL / WNE band (see Fig.~\ref{WRtype}) and the lower range of WC stars. Interestingly, the estimated initial mass of the progenitor of SN 2008ak, which exhibits a transition spectrum between type II and type Ib, is in the range of $25-30\,\mathrm{M}_\odot$ for a metallicity around solar \citep{Crockett2008a,Pastorello2008a}. This agrees with the present stellar models. Type Ic SNe cover a much broader range of initial masses than type Ib. This trend reflects that most stars which, at a given stage, in their evolution have peeled off their H-rich envelope do not stop at that stage. Their evolution drives them beyond up to the stage where most of their He-rich layers have been peeled off.

One of the main uncertainties of our models is the mass loss rate. Some recent studies suggest that this rate could be overestimated by a factor of 2 \citep{Vanbeveren2007a}. We briefly discuss here qualitatively how the results presented here would be modified if we had considered lower mass loss rates. The main differences are :
\begin{itemize}
\item The entrance in the WR phase is delayed, so this phase is shorter.
\item At a given metallicity, the minimum initial mass to obtain a WR star is increased.
\item According to this last point, the minimum initial mass providing a type Ibc SN at a given metallicity is also increased.
\end{itemize}
Globally, the decrease in the mass loss rate in our models corresponds to a shift in the limits shown in 
in Figs.~\ref{WRtype} and \ref{SNtype} up along the metallicity (vertical) axis.

\section{Nature of the stellar remnants}\label{Rema}

Superposed to Figs.~\ref{WRtype} and \ref{SNtype}, we have indicated, as a light grey zone, the regions where a black hole (BH) might be formed instead of a neutron star (NS) (see also Table~\ref{masslimits}). For drawing this zone, we assumed that $2.7\,\mathrm{M}_\odot$ is the maximum mass of a NS. This value is compatible with the one given by \citet{Shapiro1983a} and also with the recent discovery of a massive NS of $2.1\,\mathrm{M}_\odot$ \citep{Freire2008a}. By adopting this mass limit and the relation $M_{\rm NS}$ versus $M_{\rm CO}$ deduced from the models of \citet{Hirschi2005a}, we can estimate the mass ranges of stars producing a NS for each metallicity, or a BH. For $Z\le 0.01$, all stars with an initial mass higher than $30\,\mathrm{M}_\odot$ finish their life as a BH, that is, all WR stars and the most massive red-- or blue--supergiants. The minimum initial mass to become a BH is higher at higher metallicities. There is also a maximum initial mass to become a BH at high metallicity. This comes from the intense stellar winds that peel off the most massive stars and prevent them from having sufficient massive cores to produce a BH. As a result,  the mass range of BH progenitors decreases when more and more metal-rich environments are considered, and for $Z \ge \sim 0.040$, no BH is expected from the single-star scenario (see Fig.~\ref{WRtype}). 

We note that, except at low metallicity ($Z\le 0.01$), the majority of WNL and WNE stars produce a NS. WC stars may produce either NS or BH for metallicities higher than about 0.01. Below that metallicity they would only produce BH. WO stars produce only BH. In the same manner looking at Fig.~\ref{SNtype}, we see that most of type Ib's produce a NS. Type Ic's produce either a BH or a NS for metallicities above 0.02. For lower metallicities, only BH are expected to result from a type Ic event.

\section{Frequency of type Ib, Ic SNe, and comparison with observations}\label{Freq}

\begin{figure}
\centering
\includegraphics[angle=0,width=9cm]{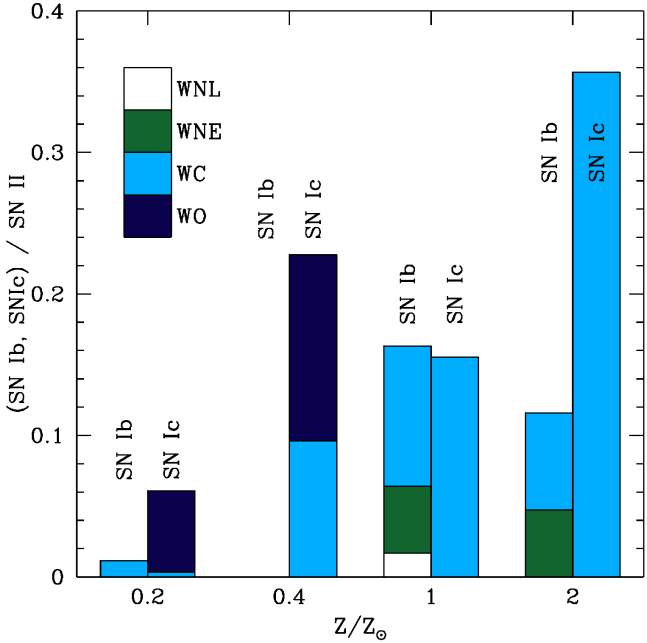}
\caption{For each metallicity, the left column represents the ratio type Ib SNe / type II SNe, and the right one the ratio type Ic / type II. The contribution of each kind of progenitor is also represented in each column. We used our reference case to built this figure (color figure available online).}
\label{SNsubtype}
\end{figure}

In Fig.~\ref{SNsubtype} we show the variation with metallicity in the number ratio of type Ib, Ic SNe for each type of progenitor with respect to type II SNe. To obtain these results, we used a Salpeter initial mass function (IMF), and apply the relation (in case of the Ib/II ratio):

\begin{equation}
\frac{\mathrm{SN\,I}_\mathrm{b}}{\mathrm{SN\,II}} = \frac{\int_{M_\mathrm{min,\, b}} ^{M_\mathrm{max,\, b}}M^{-2.35}\mathrm{d}M}{\int_8 ^{M_\mathrm{max,\, II}} M^{-2.35}\mathrm{d}M}
\end{equation}

\noindent where $M_\mathrm{min,\, b}$ is the minimum initial mass giving a type Ib SN, $M_\mathrm{max,\, b}$ the maximum initial mass giving a type Ib SN, and $M_\mathrm{max,\, II}$ the maximum initial mass giving a type II SN . We used the mass limits given in Table~\ref{masslimits}, and we supposed that an SN event occurs, even when a BH forms. Similar expressions are used to compute the other ratios with respect to type II SNe.

The first striking point when looking at Fig.~\ref{SNsubtype} is that, for most of the metallicities studied here, type Ic are more frequent than type Ib. This was already apparent in Fig.~\ref{SNtype}, but here it is confirmed after the mass ranges involved have been weighted by the IMF. Only at solar metallicities would present models predict a similar frequency between Ib and Ic. A second point deserving to be mention is that the vast majority of the type Ibc supernovae come from WC or WO stars. WN stars contribute only to a relatively small proportion of the type Ib's and only at metallicities equal to or above solar. 

\begin{figure}
\centering
\includegraphics[angle=0,width=9cm]{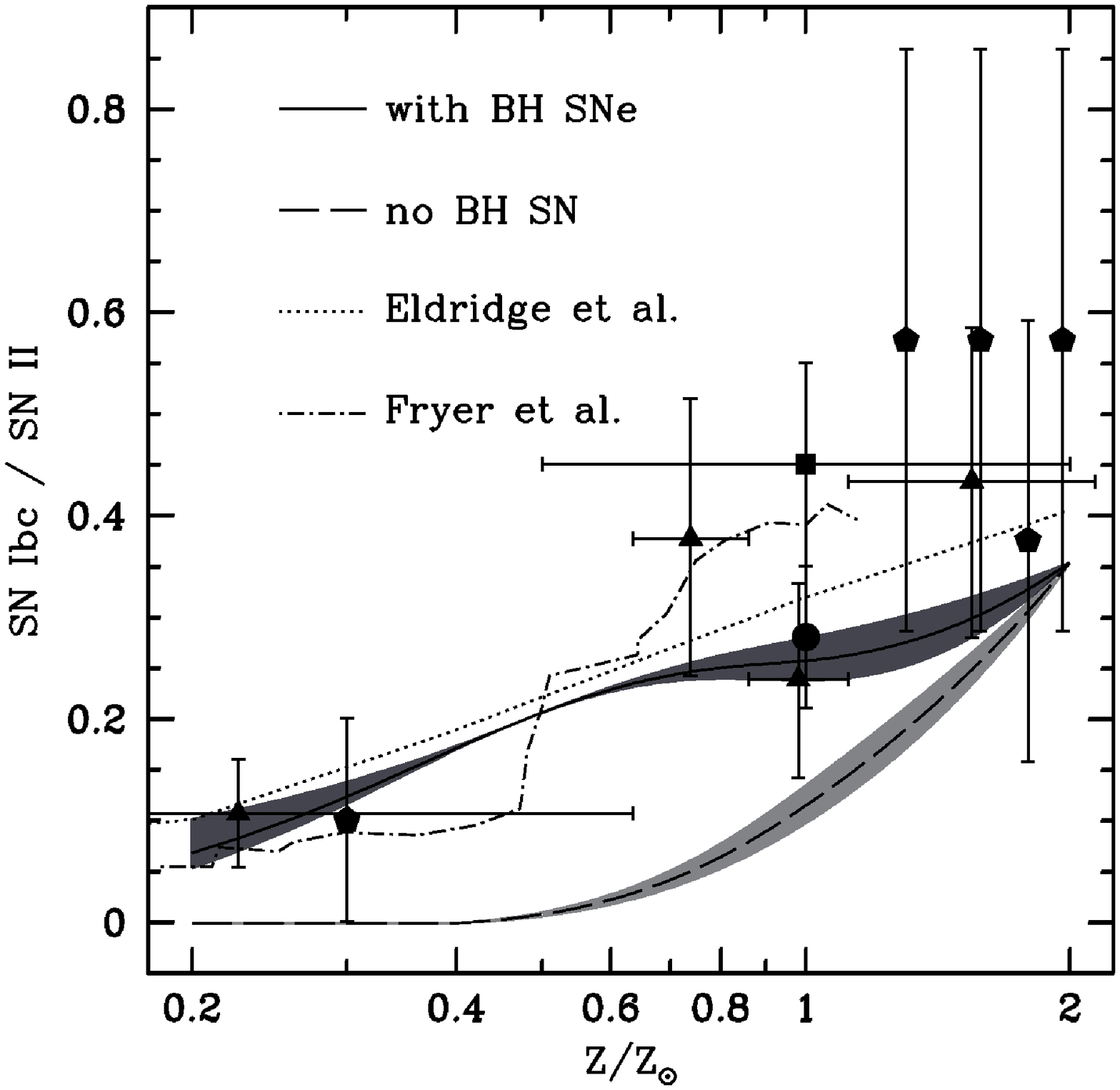}
\caption{Rate of SN Ibc / SN II if all models produce an SN (solid line) or if models producing a BH do not explode in an SN (dashed line). Grey areas are the corresponding estimated errors from our models. Pentagons are observational data from \citet{Prieto2008a}, triangles are data from \citet{Prantzos2003a}, circle, and square, are measurements at solar metallicity from \citet{Cappellaro1999a} and \citet{Smartt2009a} respectively. The dotted line represents the binary models of \citet{Eldridge2008a}, and the dotted-dashed line the rate obtained with the binary models of  \citet{Fryer2007a}.}
\label{RateBlackHole}
\end{figure}

These frequencies are compared with recent observed ratios in Fig. \ref{RateBlackHole}, and our models with the observed rates given by \citet{Cappellaro1999a,Prantzos2003a,Smartt2009a} and \citet{Prieto2008a}\footnote{In their recent paper, \citet{Prieto2008a} analyze a sample of 77 core collapse SNe (and 38 type Ia SNe) in the redshift range $0.01 < z < 0.04$. This sample is composed of 58 SNe II, 13 SNe Ic, 3 SNe Ib, and 3 SNe Ibc, \textbf{all with a measurement of  $\log(\mathrm{O}/\mathrm{H}) + 12$ for the host galaxy.}}. We also add the prediction of binary star models of \citet{Eldridge2008a} and of \citet{Fryer2007a}. The gray areas indicate the variation of our computed ratio taking the mass limits $M_\mathrm{lim}^\mathrm{II/Ib}$ obtained varying $m^\mathrm{min}_\mathrm{H}$ between 0 and 0.5. 

Our models reproduce the observations up to $Z = \mathrm{Z}_\odot$\footnote{In \citet{Meynet2005a}, we already made this comparison, although with a less sophisticated method to link the progenitor to the supernova type it produces.}. At metallicities higher than about 0.02, present models predict too few type Ibc supernovae, although in view of the big error bars, it is difficult to ascertain that this is a real deficit.  The increase in the ratio of type Ibc/type II SNe with the metallicity naturally arises from the metallicity dependence of the stellar winds. At high metallicity stellar winds are stronger, making the formation of stars without an H-rich envelope that explode as a type Ib or Ic supernova easier.

The results obtained including close binary evolutionary scenarios are also able to reproduce the general increase in the SN Ibc / SN II ratio with respect to the metallicity. In these models most of the type Ibc supernovae occur as a result of mass transfer in close binary systems. Thus we face here the situation where two very different models (single stars with rotation/close binary evolution with mass transfer) are both able to give a reasonable fit to the data. Actually both scenarios probably contribute to the observed populations of type Ibc supernovae. However, it would be interesting to know their relative importance and how their relative importance changes with the metallicity. It might be that both scenarios predict different behaviors for the way the frequencies of the type Ib and type Ic SNe vary as a function of the metallicity. Below we discuss the predictions of single-star rotating models and compare them with the available observations.

\begin{figure}
\centering
\includegraphics[angle=0,width=9cm]{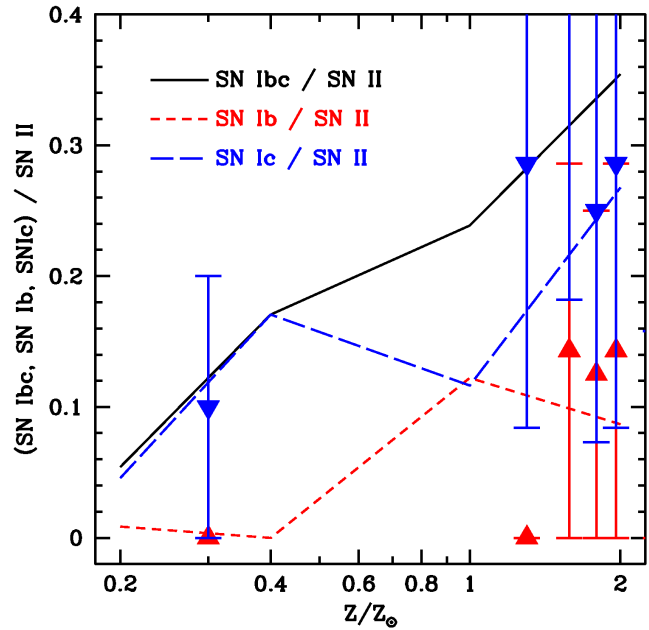}
\caption{Rates of SN Ic / SN II (blue long--dashed line), SN Ib / SN II (red short--dashed line), and SN Ibc / SN II (black solid line) for the reference case. The points are extracted from the data of \citet{Prieto2008a}: triangles (red) represent the SN Ib / SN II rate, and upside--down triangles (blue) the SN Ic / SN II rate. Each triangle corresponds to a sample of 11 core collapse SNe (color figure available online).}
\label{SNRatesHighHe}
\end{figure}

From the data of \citet{Prieto2008a}, we extract separate observed rates for type Ib and type Ic SNe. They are represented for the SN Ib / SN II rate, and for SN Ic / SN II in Fig.~\ref{SNRatesHighHe}. We note that, in the high metallicity range, the observed rate of type Ib SNe is lower than the rate of type Ic by about a factor two. The trend deduced from the present theoretical models is in good agreement with the observed one.

\begin{figure}
\centering
\includegraphics[angle=0,width=9cm]{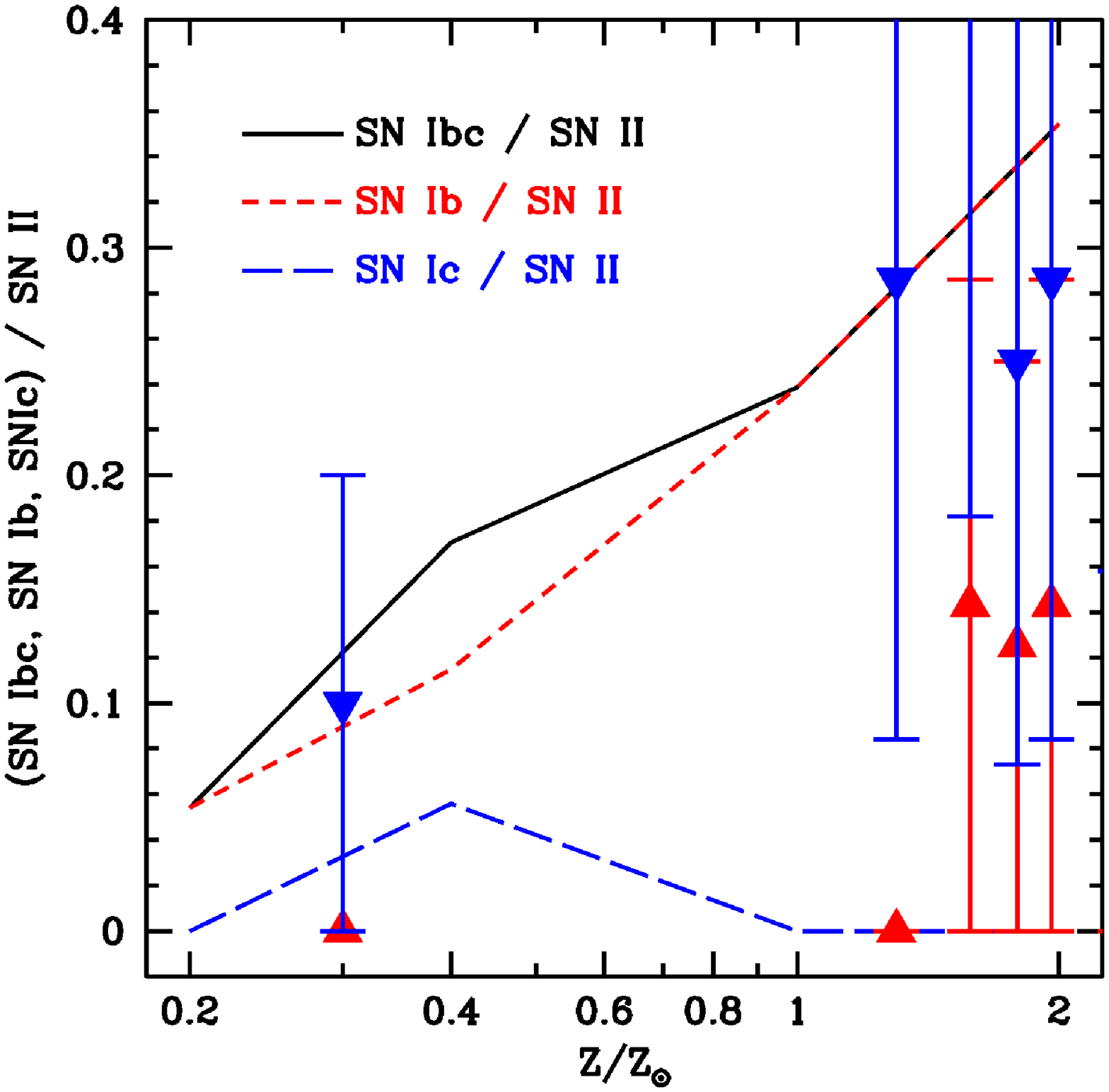}
\caption{Same as Fig.~\ref{SNRatesHighHe} but with $m^\mathrm{min}_\mathrm{He}=0.4\,\mathrm{M}_\odot$ instead of $0.6$ (see text).}
\label{SNRatesLowHe}
\end{figure}

If we adopt $m^\mathrm{min}_\mathrm{He}=0.4$ instead of 0.6 (see Fig.~\ref{SNRatesLowHe}), we see that our models never produce type Ic SNe (except a small number at $Z = 0.4\,\mathrm{Z}_\odot$). In that case, models would predict that most type Ibc supernovae would be type Ib, which is contrary to what is observed. This numerical experiment illustrates the strong dependence of the present results on the maximum quantity of helium in the ejecta that can be accommodated in a type Ic supernova event. In case it were confirmed that this value is as low as $0.4\,\mathrm{M}_\odot$, then present models would not provide a good fit to the observed data. Enhancing this quantity to values superior or equal to 0.55-0.6 suffices to provide a good fit. 

\begin{figure}
\centering
\includegraphics[angle=0,width=9cm]{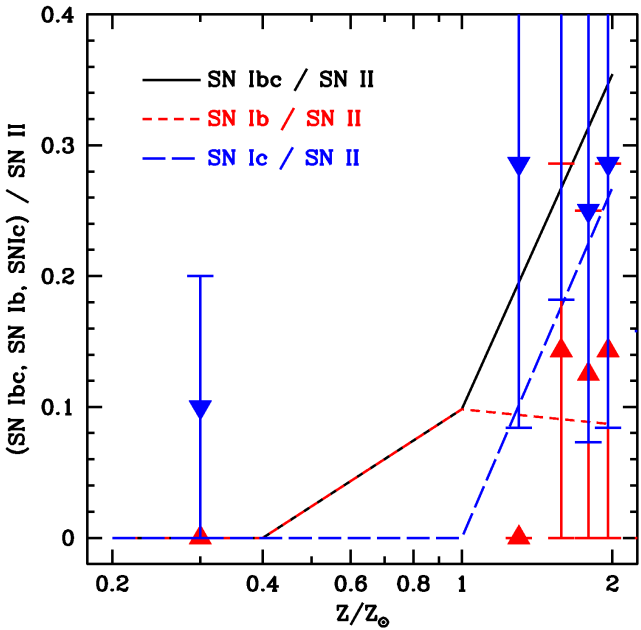}
\caption{Same as Fig. \ref{SNRatesHighHe}, with the assumption that when a black hole is formed, no SN event occurs (color figure available online).}
\label{SNRateHighHeBH}
\end{figure}

How do the results change when one makes the hypothesis that when a BH is formed no SN event occurs? The situation for the variation with metallicity in the number ratio of Type Ibc to type II SNe is shown in Fig.~\ref{RateBlackHole}. In Fig.~\ref{SNRateHighHeBH}, we show what would be the predictions of the present models for the frequencies of type Ib and type Ic taken separately and adopting $m^\mathrm{min}_\mathrm{He}=0.6$. At low metallicity, no type Ibc SNe are expected. At $Z \sim 0.010 = 0.5\,\mathrm{Z}_\odot$, type Ib SNe begin to come out of the BH domain, making the SN Ib / SN II ratio begins to increase.  For  $Z \ge 0.02$,  single-star models might still account for more than half of the type Ibc supernovae. However, at $Z=0.02$,  all the SNe coming from single-star evolution would have type Ib, which is not  consistent with the observations. At twice solar metallicity, all our models lead to a neutron star, and the ratios are identical to the one obtained in Fig.~6.  These comparisons show that, considering that the hypothesis ``when a BH is formed no SN event occurs'' makes a big difference for metallicities below about 0.02. Above that metallicity, the results are only weakly affected.

If we take a lower maximal mass for neutron stars, this situation becomes more extreme: almost all the WR stars produce a BH, thus the rate of SN Ibc / SN II is zero or very low at every metallicity. This illustrates the high sensitivity of the above results on either the mass limit for forming BH and/or the possibility or impossibility to have a SN event when a BH is formed.

Probably, the hypothesis according to which no supernova event appears when a BH is formed is too restrictive. Until no clear understanding of the physics involved in core collapse supernova explosion is reached, it is difficult to draw firm conclusions at that point, especially when one considers the possible effects of rotation. In that context, the collapsar scenario for GRBs \citep{Woosley1993a} needs the formation of a BH \citep{Dessart2008a}, and this formation is at least accompanied in some cases by a type Ic supernova event. Also, the observation of  the binary system GRO J1655-40 containing a BH \citep{Israelian1999a} suggests that a few stellar masses have been ejected when the black hole formed and therefore that an SN event occurred. This is suggested by the important chemical anomalies observed at the surface of the companion whose origin is attributed the companion being accreted a part of the SN ejecta. These examples show that there are at least some circumstances where the formation of a BH does not prevent an SN event to occur. Probably, reality is somewhere between the two extreme cases discussed before: ``all massive stars produce an SN, independently of the remnant type", and ``only the stars leading to a neutron star produce an SN".

As in Sect.~\ref{SN}, it is interesting here to infer how a decrease in the mass loss rate modifies the previous results. At first order, a decrease in the mass loss rates is equivalent to a shift in the curves presented in Figs.~\ref{RateBlackHole} to \ref{SNRateHighHeBH} to the right along the metallicity (horizontal) axis. In general, in that case, the theoretical ratios will be lower than the observed values.

\section{Gamma ray burst progenitors}\label{GRB}

It is now admitted that there is a connection between so-called long-soft GRB events and type Ic supernovae \citep{Woosley2006a}. As seen in the previous section, this kind of supernova can be produced by both WC or WO stars. In this way, WC and WO stars are natural candidates to be GRB progenitors. However, GRB are primarily found in metal--poor environments \citep{Modjaz2008a}, while type Ic's SNe appear mainly at high metallicity.  Many physical reasons have been invoked to explain why GRBs seem to occur only at low metallicities. Among them are the following \citep[see the review by][]{Woosley2006a}:
\begin{itemize}
\item At low metallicity, stellar winds (even during the WR phases) are weaker, thus bringing away small quantities of angular momentum. Black holes are also more easily formed since higher final masses are obtained.
\item At low metallicity the transport of angular momentum between the core and the envelope is less efficient than at high metallicities, because of slower meridional currents in metal poor stars.
\item Since the chemical mixing due to rotation is more efficient at low $Z$, homogeneous evolution is more easily obtained in metal-poor regions. Homogeneous evolution allows massive stars to produce a type Ic SN event without having to lose large amounts of mass. Indeed, a perfectly homogeneous evolution (actually never realized) would allow the formation of a pure CO core (and then lead to a type Ic SN event) at the end of the core He-burning phase without the need for the star to lose any mass!
\item The distribution of initial velocities at low metallicity might contain more fast rotators than at high metallicities \citep[see Fig. 9 in ][]{Martayan2007a}.
\end{itemize}
At the moment, these arguments remain quite speculative.

\begin{figure}
\centering
\includegraphics[angle=0,width=9cm]{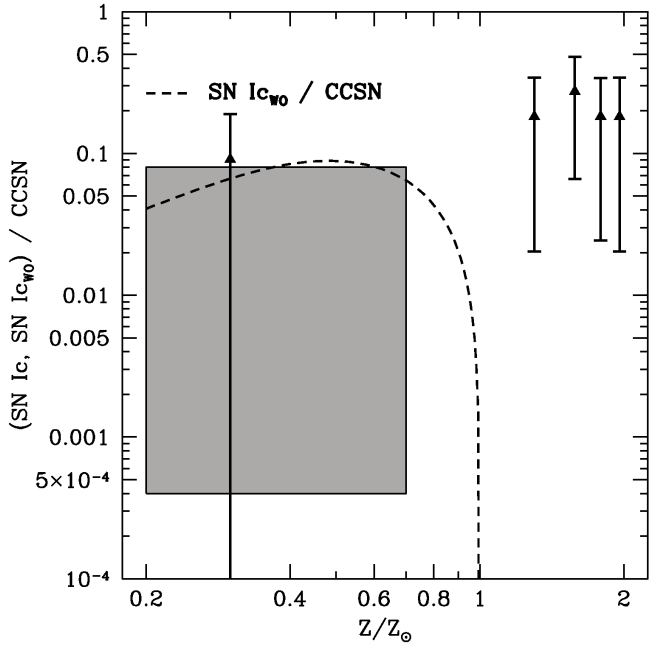}
\caption{Observed rate of type Ic SNe (triangles with error bars) and predicted rate of type Ic SNe whose progenitor is a WO star (dashed line) with respect to the total number of core collapse SNe (CCSNe). The gray rectangle represents the extension in metallicity \citep{Modjaz2008a} and rate \citep{Podsiadlowski2004a} of GRB events.}
\label{GRBRate}
\end{figure}

What can be done presently is to compare the observed GRB frequency with the observed frequency of  potential candidates. First, as mentioned above, type Ic supernovae at low metallicity do appear interesting candidates. From Fig.~\ref{GRBRate}, one can see that the observed rate of type Ic supernovae from single star models is still above the estimated number ratio GRB / core collapse supernovae (CCSNe) even when one only considers low-metallicity, type Ic SNe. Here we supposed that the formation of a BH does not prevent an SN event. In case of course only rare circumstances would allow an SN to occur when a black hole is formed, then the situation might be very different (see below).

Other interesting candidates are the WO stars that primarily occur at low metallicity. Table~\ref{WOCat} lists all the known WO stars. If we take the value $\log(\mathrm{O} / \mathrm{H}) + 12 = 8.66$ \citep{Grevesse2007a} for the solar oxygen abundance, we see that  6 out of a total of 8 occur in regions with a metallicity under $Z/\mathrm{Z}_\odot < 0.9$. These stars will explode as a type Ic SN. The frequency of type Ic's SNe with WO star progenitors is shown in Fig.~\ref{GRBRate}. The expected rate is only marginally compatible with the observed GRB rate (assuming that the aperture angle of the bipolar jet is very small, typically around $1^\mathrm{o}$). This conclusion has already been obtained by \citet{Hirschi2005a}, so even restraining the progenitors of GRB to WO stars would still not match the observed frequency of GRBs. 

\begin{table*}
\caption{WO stars catalog.}
\label{WOCat}
\begin{center}
\begin{tabular}{c c c c c c c}
\hline\hline
Name & Galaxy & galactocentric radius & $\log(\mathrm{O}/\mathrm{H)} + 12$ & Cat number & type & Ref \\
\hline\\
DBU 1 & MW & 4.6 & 8.76\textsuperscript{2} & WR 93b & WO3 & 3, 4\\
MS 4 & MW & 9.2 & 8.56\textsuperscript{2} & WR 30a & WO3 & 1\\
Sand 4 & MW & 3.4 & 8.81\textsuperscript{2} & WR 102 & WO2 & 1, 3\\
Sand 5 & MW & 7.8 & 8.62\textsuperscript{2} & WR 142 & WO2 & 1\\
Sand 1 & SMC & & 8.05 & & WO4 & 5, 6\\
Sand 2 & LMC & & 8.35 & & WO3 & 5, 3\\
DR 1 & IC 1613 & & 7.90 & & WO3 & 7, 8\\
NGC 1313 \#31 & NGC 1313 & & 8.23 & & WO3 & 9\\
\hline
\multicolumn{7}{l}{(1) \citet{VanDerHucht2001a}, (2) \citet{Rood2007a}, (3) \citet{Drew2004a}, (4) \citet{VanDerHucht2006a},}\\
\multicolumn{7}{l}{(5) \citet{Hunter2007a}, (6) \citet{StLouis2005a}, (7) \citet{Kingsburgh1995a},}\\
\multicolumn{7}{l}{(8) \citet{Bresolin2007a}, (9) \citet{Hadfield2007a}}
\end{tabular}
\end{center}
Note: 1st column is the name, 2nd one is the host galaxy, 3rd one is the galactocentric radius (only for Galactic WO), 4th one is the oxygen abundance, 5th one is the number in \textit{The VIIth Catalogue of Galactic Wolf-Rayet Stars}\textsuperscript{1, 4}, 6th column is the type of the WO star, and the last one the references.
\end{table*}

From the above discussion, it can be deduced that exploding as a type Ic SN in metal-poor regions (or having a WO progenitor) is not a sufficient condition for a GRB. Physical characteristics shared by a subsample of the metal poor type Ic events exists that are needed to obtain a GRB event. Probably this physical characteristic is the high angular momentum in the core (Yoon et al. 2006; Woosley \& Heger 2006). 

If GRB would only occur for initially very fast-rotating stars, rotating so fast that these stars would follow a homogeneous evolution \citep{Maeder1987a,Yoon2006a,Woosley2006b,Meynet2007a}, can we expect to find any peculiar feature in the chemical composition of the ejecta testifying this previous homogeneous evolution? Or in other words, is there any difference in the chemical composition of the ejecta between a type Ic having ``normal WC or WO" progenitors and those arising from a model that followed a homogeneous evolution during the MS phase? To that purpose we computed the evolution of a $60\,\mathrm{M}_\odot$ model followed up to the end of central Si--burning, with an initial metallicity of $Z=0.002$ and an initial rotation rate of $\Omega / \Omega_\mathrm{crit} = 0.75$. The model was computed including magnetic field, following \citet{Spruit2002a} and \citet{Maeder2005a}. This model follows a nearly homogeneous evolutionary track during the MS phase. The angular momentum content at the time of the presupernova is sufficient for producing a collapsar. In Tables~\ref{HeMass} and \ref{EjMass}, we present various characteristics of that model.

Table~\ref{HeMass} shows that the model is in the WC phase at the end of its evolution. The mixing induced by the magnetic field produces larger cores than in the models without magnetic field. Thus, due to mass loss, the core is uncovered at an early stage of He--burning, leading to a WC star. The helium content of the ejecta ($0.64\,\mathrm{M}_\odot$) is also slightly greater than in our other models of similar mass and is also just above the value of $0.6\,\mathrm{M}_\odot$ below which a type Ic is supposed to occur in our classification scheme. This implies that our homogenous progenitors lead to a type Ib SN. On the other hand, as mentioned above, the limit of 0.6 has not to be taken as a firm limit, and stars with higher He-content might also explode in a type Ic event.  Moreover, the models having followed a homogeneous evolution will end their evolution with more angular momentum in their central regions. This may have a significant impact on the way the stars explode, as well as on the spectral features arising from the SN explosion!

It is also interesting to compare the chemical composition of the ejecta as they are obtained in the present models when they follow a homogeneous evolution or not. Compare for instance the 60 M$_\odot$ at $Z=0.004$ (heterogeneous case) with the 60 M$_\odot$ at $Z=0.002$ (homogeneous case). The differences in the masses of He, CNO elements, and $Z$ are very small. Thus there is little chance from the observations of the composition of the ejecta in CNO elements to be able to distinguish between a normal and a homogeneous evolution.

\begin{figure}
\centering
\includegraphics[angle=0,width=9cm]{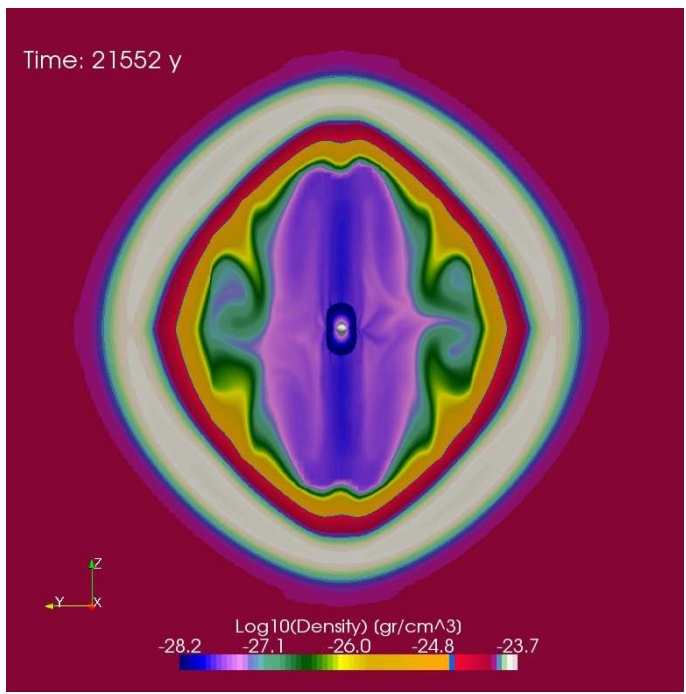}
\caption{Young bi-modal shaped nebula blown from the highly aspherical, polar dominated winds around a $20\,\mathrm{M}_\odot$ star at $Z=10^{-5}$. The star has evolved during $21552\,\mathrm{yr}$ since it has left the ZAMS. The maximal extension of the wind is $\sim1.5\cdot 10^{18}\mathrm{cm}$. This model was computed by the A-Maze code \citep{Walder2000a}  (color figure available online).}
\label{PolWind}
\end{figure}

A point that would be interesting to check in a later work is the following: since the star rotate very fast during a great part of its evolution, it will produce anisotropic stellar winds (see Fig. \ref{PolWind}), typically bipolar winds and probably equatorial mass loss when the critical limit is reached \citep{Maeder1999a}. These anisotropic winds will shape the circumstellar environment of the star \citep[][and see also Fig.~\ref{PolWind}]{vanMarle2008a} and it might be that some traces of the resulting particular morphology will still be present at the time of the SN event (for instance, equatorial mass loss probably gives rise to a slow equatorial expanding disk whose traces might still be present when the star explodes). In that case the circumstellar environment of GRB may be peculiar. This can in turn have an impact on some features in the spectrum. More generally, the circumstellar environment of stars that have been in a not too remote past very fast rotators may also be characterized by such features. It would of course be very interesting to find in the circumstellar environment of some stars, now slowly rotating, traces of their very rapid rotation in a previous phase of their evolution.

\section{Conclusion}\label{Conc}

In \citet{Meynet2005a} we showed that single-star models can account for the increase with the metallicity in the ratio (SN Ib+SN Ic)/SNII. In the present work we obtained the following additional results:
\begin{itemize}
\item Single-star models indicate that the mass of helium ejected in type Ic supernovae is at least 0.3 M$_\odot$ \citep[see also][]{Eldridge2004a}. 
\item Type Ib supernovae may be produced by WNL, WNE, or less massive WC stars. Type Ic SNe are only produced by WC or WO stars.
\item If we consider that the ejecta of a type Ic supernova may contain up to $0.6\,\mathrm{M}_\odot$ of He and that an SN event occurs even when a BH is formed, single-star models are able to reproduce the observational cumulative ratio of SN Ibc / SN II well along with both detailed ratios SN Ib / SN II and SN Ic / SN II. Because of the small number of observed events with known $\log(\mathrm{O}/\mathrm{H})$, the actual estimations of the ratios SN Ib/SN II and of SN Ic/SN II have large uncertainties. Additional measurements  will permit better constrained stellar models. 
\item In the framework of the ``BH-no supernova event" hypothesis, all type Ibc supernovae at low metallicity should come from close binary evolution.  At solar and higher-than-solar metallicities, more than half of the type Ibc might still be explained by single-star models.
\item Our models predict too high a rate of GRB, even considering that only WO stars lead to this rare event. Probably GRB only originate in the most rapid rotators.
\item There is no way to distinguish on the basis of the chemical composition of the ejecta a type Ic supernova arising from a normal heterogeneous evolution from a type Ic supernova arising from a homogeneous evolution. The past fast rotation of the GRB progenitor may have left some imprint on the circumstellar matter distribution.
\item For $Z \ge 0.04$, all massive stars produce neutron stars. When the metallicity decreases, the range of initial mass producing neutron stars decreases. Below a metallicity of about $0.01$, all stars more massive than about $30\,\mathrm{M}_\odot$ produce a black hole.
\end{itemize}

It does appear difficult on the basis of the variation in the ratio (SN Ib + SN Ic)/SN II with the metallicity to differentiate the relative importance of the single and binary channels. As can be seen from Fig.~\ref{RateBlackHole}, scenarios with both a significant proportion of close binaries and with a modest one are able to reproduce the observational trend. To make progress it would be interesting to study the predictions of the close binary channel for the relative frequency of type Ib and type Ic supernovae and to compare them with the observations. Also it would be useful  to have predictions similar to those presented here for what concerns the chemical composition of the ejecta. 

One can also note that, in case the hypothesis ``no SN occurs when a BH forms" is correct, the binary scenario is responsible for most, if not all, type Ic and type Ib events at low metallicities, while at high metallicities single stars would still account for more than one half of the type Ibc supernovae. In the framework of that hypothesis, the long-soft GRB (at least those associated to type Ic events) would originate in binary systems. Let us note however that some observations show that at least in some circumstances, a SN occurs when a BH is formed.

A promising way to study the respective importance of the single and close binary channel for the formation of type Ibc SN progenitors is to look for single-aged stellar populations with masses at the turn off between $8$ and $25\,\mathrm{M}_\odot$. If such clusters very rarely show both RSG and WR stars, then this can be interpreted as these two types of stars coming from very different mass ranges just as would be expected in single-star scenarios. If on the contrary, the simultaneous occurrence of both RSG and WR stars is frequent, then it may indicate that there is an overlap of the initial mass ranges of progenitors of RSG and WR stars. In that case, close binary evolution is probably required (except maybe at high metallicity where the overlap also exists for single-star scenarios).

\vspace{1 cm}
\textit{Acknowledgments} The authors would like to thank the referee Philippe Podsiadlowski, as well as John J. Eldridge and Stephen J. Smartt for their precious comments that improved this work.

\bibliographystyle{aa}
\bibliography{MyBiblio}

\begin{thebibliography}{62}
\expandafter\ifx\csname natexlab\endcsname\relax\def\natexlab#1{#1}\fi

\bibitem[{{Bresolin} {et~al.}(2007){Bresolin}, {Urbaneja}, {Gieren},
  {Pietrzy{\'n}ski}, \& {Kudritzki}}]{Bresolin2007a}
{Bresolin}, F., {Urbaneja}, M.~A., {Gieren}, W., {Pietrzy{\'n}ski}, G., \&
  {Kudritzki}, R.-P. 2007, \apj, 671, 2028

\bibitem[{{Cappellaro} {et~al.}(1999){Cappellaro}, {Evans}, \&
  {Turatto}}]{Cappellaro1999a}
{Cappellaro}, E., {Evans}, R., \& {Turatto}, M. 1999, \aap, 351, 459

\bibitem[{{Crockett} {et~al.}(2008){Crockett}, {Eldridge}, {Smartt},
  {Pastorello}, {Gal-Yam}, {Fox}, {Leonard}, {Kasliwal}, {Mattila}, {Maund},
  {Stephens}, \& {Danziger}}]{Crockett2008a}
{Crockett}, R.~M., {Eldridge}, J.~J., {Smartt}, S.~J., {et~al.} 2008, \mnras,
  391, L5

\bibitem[{{de Jager} {et~al.}(1988){de Jager}, {Nieuwenhuijzen}, \& {van der
  Hucht}}]{deJager1988a}
{de Jager}, C., {Nieuwenhuijzen}, H., \& {van der Hucht}, K.~A. 1988, \aaps,
  72, 259

\bibitem[{{Dessart} {et~al.}(2008){Dessart}, {Burrows}, {Livne}, \&
  {Ott}}]{Dessart2008a}
{Dessart}, L., {Burrows}, A., {Livne}, E., \& {Ott}, C.~D. 2008, \apjl, 673,
  L43

\bibitem[{{Drew} {et~al.}(2004){Drew}, {Barlow}, {Unruh}, {Parker}, {Wesson},
  {Pierce}, {Masheder}, \& {Phillipps}}]{Drew2004a}
{Drew}, J.~E., {Barlow}, M.~J., {Unruh}, Y.~C., {et~al.} 2004, \mnras, 351, 206

\bibitem[{{Dufton} {et~al.}(2006){Dufton}, {Smartt}, {Lee}, {Ryans}, {Hunter},
  {Evans}, {Herrero}, {Trundle}, {Lennon}, {Irwin}, \& {Kaufer}}]{Dufton2006a}
{Dufton}, P.~L., {Smartt}, S.~J., {Lee}, J.~K., {et~al.} 2006, \aap, 457, 265

\bibitem[{{Ekstr{\"o}m} {et~al.}(2008){Ekstr{\"o}m}, {Meynet}, {Chiappini},
  {Hirschi}, \& {Maeder}}]{Ekstrom2008a}
{Ekstr{\"o}m}, S., {Meynet}, G., {Chiappini}, C., {Hirschi}, R., \& {Maeder},
  A. 2008, \aap, 489, 685

\bibitem[{{Eldridge} {et~al.}(2008){Eldridge}, {Izzard}, \&
  {Tout}}]{Eldridge2008a}
{Eldridge}, J.~J., {Izzard}, R.~G., \& {Tout}, C.~A. 2008, \mnras, 384, 1109

\bibitem[{{Eldridge} \& {Tout}(2004)}]{Eldridge2004a}
{Eldridge}, J.~J. \& {Tout}, C.~A. 2004, \mnras, 353, 87

\bibitem[{{Elmhamdi} {et~al.}(2006){Elmhamdi}, {Danziger}, {Branch},
  {Leibundgut}, {Baron}, \& {Kirshner}}]{Elmhamdi2006a}
{Elmhamdi}, A., {Danziger}, I.~J., {Branch}, D., {et~al.} 2006, \aap, 450, 305

\bibitem[{{Freire} {et~al.}(2008){Freire}, {Wolszczan}, {van den Berg}, \&
  {Hessels}}]{Freire2008a}
{Freire}, P.~C.~C., {Wolszczan}, A., {van den Berg}, M., \& {Hessels}, J.~W.~T.
  2008, \apj, 679, 1433

\bibitem[{{Fryer} {et~al.}(2007){Fryer}, {Mazzali}, {Prochaska}, {Cappellaro},
  {Panaitescu}, {Berger}, {van Putten}, {van den Heuvel}, {Young},
  {Hungerford}, {Rockefeller}, {Yoon}, {Podsiadlowski}, {Nomoto}, {Chevalier},
  {Schmidt}, \& {Kulkarni}}]{Fryer2007a}
{Fryer}, C.~L., {Mazzali}, P.~A., {Prochaska}, J., {et~al.} 2007, \pasp, 119,
  1211

\bibitem[{{Grevesse} {et~al.}(2007){Grevesse}, {Asplund}, \&
  {Sauval}}]{Grevesse2007a}
{Grevesse}, N., {Asplund}, M., \& {Sauval}, A.~J. 2007, Space Science Reviews,
  130, 105

\bibitem[{{Hadfield} \& {Crowther}(2007)}]{Hadfield2007a}
{Hadfield}, L.~J. \& {Crowther}, P.~A. 2007, \mnras, 381, 418

\bibitem[{{Hirschi} {et~al.}(2004){Hirschi}, {Meynet}, \&
  {Maeder}}]{Hirschi2004a}
{Hirschi}, R., {Meynet}, G., \& {Maeder}, A. 2004, \aap, 425, 649

\bibitem[{{Hirschi} {et~al.}(2005){Hirschi}, {Meynet}, \&
  {Maeder}}]{Hirschi2005a}
{Hirschi}, R., {Meynet}, G., \& {Maeder}, A. 2005, \aap, 443, 581

\bibitem[{{Huang} \& {Gies}(2006)}]{Huang2006a}
{Huang}, W. \& {Gies}, D.~R. 2006, \apj, 648, 580

\bibitem[{{Hunter} {et~al.}(2007){Hunter}, {Dufton}, {Smartt}, {Ryans},
  {Evans}, {Lennon}, {Trundle}, {Hubeny}, \& {Lanz}}]{Hunter2007a}
{Hunter}, I., {Dufton}, P.~L., {Smartt}, S.~J., {et~al.} 2007, \aap, 466, 277

\bibitem[{{Israelian} {et~al.}(1999){Israelian}, {Rebolo}, {Basri}, {Casares},
  \& {Mart{\'{\i}}n}}]{Israelian1999a}
{Israelian}, G., {Rebolo}, R., {Basri}, G., {Casares}, J., \& {Mart{\'{\i}}n},
  E.~L. 1999, \nat, 401, 142

\bibitem[{{Kingsburgh} \& {Barlow}(1995)}]{Kingsburgh1995a}
{Kingsburgh}, R.~L. \& {Barlow}, M.~J. 1995, \aap, 295, 171

\bibitem[{{Kudritzki} \& {Puls}(2000)}]{Kudritzki2000a}
{Kudritzki}, R.-P. \& {Puls}, J. 2000, \araa, 38, 613

\bibitem[{{Leitherer}(1999)}]{Leitherer1999a}
{Leitherer}, C. 1999, in IAU Symposium, Vol. 193, Wolf-Rayet Phenomena in
  Massive Stars and Starburst Galaxies, ed. K.~A. {van der Hucht},
  G.~{Koenigsberger}, \& P.~R.~J. {Eenens}, 526--+

\bibitem[{{Maeder}(1987)}]{Maeder1987a}
{Maeder}, A. 1987, \aap, 178, 159

\bibitem[{{Maeder}(1992)}]{Maeder1992a}
{Maeder}, A. 1992, \aap, 264, 105

\bibitem[{{Maeder}(1999)}]{Maeder1999a}
{Maeder}, A. 1999, \aap, 347, 185

\bibitem[{{Maeder} \& {Meynet}(2000)}]{Maeder2000a}
{Maeder}, A. \& {Meynet}, G. 2000, \aap, 361, 159

\bibitem[{{Maeder} \& {Meynet}(2001)}]{Maeder2001a}
{Maeder}, A. \& {Meynet}, G. 2001, \aap, 373, 555

\bibitem[{{Maeder} \& {Meynet}(2005)}]{Maeder2005a}
{Maeder}, A. \& {Meynet}, G. 2005, \aap, 440, 1041

\bibitem[{{Maeder} {et~al.}(2008){Maeder}, {Meynet}, {Ekstr{\"o}m}, {Hirschi},
  \& {Georgy}}]{Maeder2008a}
{Maeder}, A., {Meynet}, G., {Ekstr{\"o}m}, S., {Hirschi}, R., \& {Georgy}, C.
  2008, in IAU Symposium, Vol. 250, IAU Symposium, 3--16

\bibitem[{{Martayan} {et~al.}(2007){Martayan}, {Fr{\'e}mat}, {Hubert},
  {Floquet}, {Zorec}, \& {Neiner}}]{Martayan2007a}
{Martayan}, C., {Fr{\'e}mat}, Y., {Hubert}, A.-M., {et~al.} 2007, \aap, 462,
  683

\bibitem[{{Meynet} \& {Maeder}(2000)}]{Meynet2000a}
{Meynet}, G. \& {Maeder}, A. 2000, \aap, 361, 101

\bibitem[{{Meynet} \& {Maeder}(2003)}]{Meynet2003a}
{Meynet}, G. \& {Maeder}, A. 2003, \aap, 404, 975, paper X

\bibitem[{{Meynet} \& {Maeder}(2005)}]{Meynet2005a}
{Meynet}, G. \& {Maeder}, A. 2005, \aap, 429, 581, paperXI

\bibitem[{{Meynet} \& {Maeder}(2007)}]{Meynet2007a}
{Meynet}, G. \& {Maeder}, A. 2007, \aap, 464, L11

\bibitem[{{Meynet} {et~al.}(1994){Meynet}, {Maeder}, {Schaller}, {Schaerer}, \&
  {Charbonnel}}]{Meynet1994a}
{Meynet}, G., {Maeder}, A., {Schaller}, G., {Schaerer}, D., \& {Charbonnel}, C.
  1994, \aaps, 103, 97

\bibitem[{{Modjaz} {et~al.}(2008){Modjaz}, {Kewley}, {Kirshner}, {Stanek},
  {Challis}, {Garnavich}, {Greene}, {Kelly}, \& {Prieto}}]{Modjaz2008a}
{Modjaz}, M., {Kewley}, L., {Kirshner}, R.~P., {et~al.} 2008, \aj, 135, 1136

\bibitem[{{Nomoto} {et~al.}(1994){Nomoto}, {Yamaoka}, {Pols}, {van den Heuvel},
  {Iwamoto}, {Kumagai}, \& {Shigeyama}}]{Nomoto1994a}
{Nomoto}, K., {Yamaoka}, H., {Pols}, O.~R., {et~al.} 1994, \nat, 371, 227

\bibitem[{{Nugis} \& {Lamers}(2000)}]{Nugis2000a}
{Nugis}, T. \& {Lamers}, H.~J.~G.~L.~M. 2000, \aap, 360, 227

\bibitem[{{Pastorello} {et~al.}(2008){Pastorello}, {Kasliwal}, {Crockett},
  {Valenti}, {Arbour}, {Itagaki}, {Kaspi}, {Gal-Yam}, {Smartt}, {Griffith},
  {Maguire}, {Ofek}, {Seymour}, {Stern}, \& {Wiethoff}}]{Pastorello2008a}
{Pastorello}, A., {Kasliwal}, M.~M., {Crockett}, R.~M., {et~al.} 2008, \mnras,
  389, 955

\bibitem[{{Podsiadlowski} {et~al.}(1992){Podsiadlowski}, {Joss}, \&
  {Hsu}}]{Podsiadlowski1992a}
{Podsiadlowski}, P., {Joss}, P.~C., \& {Hsu}, J.~J.~L. 1992, \apj, 391, 246

\bibitem[{{Podsiadlowski} {et~al.}(2004){Podsiadlowski}, {Mazzali}, {Nomoto},
  {Lazzati}, \& {Cappellaro}}]{Podsiadlowski2004a}
{Podsiadlowski}, P., {Mazzali}, P.~A., {Nomoto}, K., {Lazzati}, D., \&
  {Cappellaro}, E. 2004, \apjl, 607, L17

\bibitem[{{Prantzos} \& {Boissier}(2003)}]{Prantzos2003a}
{Prantzos}, N. \& {Boissier}, S. 2003, \aap, 406, 259

\bibitem[{{Prieto} {et~al.}(2008){Prieto}, {Stanek}, \& {Beacom}}]{Prieto2008a}
{Prieto}, J.~L., {Stanek}, K.~Z., \& {Beacom}, J.~F. 2008, \apj, 673, 999

\bibitem[{{Rood} {et~al.}(2007){Rood}, {Quireza}, {Bania}, {Balser}, \&
  {Maciel}}]{Rood2007a}
{Rood}, R.~T., {Quireza}, C., {Bania}, T.~M., {Balser}, D.~S., \& {Maciel},
  W.~J. 2007, in Astronomical Society of the Pacific Conference Series, Vol.
  374, From Stars to Galaxies: Building the Pieces to Build Up the Universe,
  ed. A.~{Vallenari}, R.~{Tantalo}, L.~{Portinari}, \& A.~{Moretti}, 169--+

\bibitem[{{Shapiro} \& {Teukolsky}(1983)}]{Shapiro1983a}
{Shapiro}, S.~L. \& {Teukolsky}, S.~A. 1983, {Black holes, white dwarfs, and
  neutron stars: The physics of compact objects} (Research supported by the
  National Science Foundation.~New York, Wiley-Interscience, 1983, 663 p.), 264

\bibitem[{{Smartt} {et~al.}(2009){Smartt}, {Eldridge}, {Crockett}, \&
  {Maund}}]{Smartt2009a}
{Smartt}, S.~J., {Eldridge}, J.~J., {Crockett}, R.~M., \& {Maund}, J.~R. 2009,
  \mnras, 508

\bibitem[{{Smith} \& {Maeder}(1991)}]{Smith1991a}
{Smith}, L.~F. \& {Maeder}, A. 1991, \aap, 241, 77

\bibitem[{{Spruit}(2002)}]{Spruit2002a}
{Spruit}, H.~C. 2002, \aap, 381, 923

\bibitem[{{St-Louis} {et~al.}(2005){St-Louis}, {Moffat}, {Marchenko}, \&
  {Pittard}}]{StLouis2005a}
{St-Louis}, N., {Moffat}, A.~F.~J., {Marchenko}, S., \& {Pittard}, J.~M. 2005,
  \apj, 628, 953

\bibitem[{{van der Hucht}(2001)}]{VanDerHucht2001a}
{van der Hucht}, K.~A. 2001, VizieR Online Data Catalog, 3215, 0

\bibitem[{{van der Hucht}(2006)}]{VanDerHucht2006a}
{van der Hucht}, K.~A. 2006, \aap, 458, 453

\bibitem[{{van Marle} {et~al.}(2008){van Marle}, {Langer}, {Yoon}, \&
  {Garc{\'{\i}}a-Segura}}]{vanMarle2008a}
{van Marle}, A.~J., {Langer}, N., {Yoon}, S.-C., \& {Garc{\'{\i}}a-Segura}, G.
  2008, \aap, 478, 769

\bibitem[{{Vanbeveren} {et~al.}(2007){Vanbeveren}, {Van Bever}, \&
  {Belkus}}]{Vanbeveren2007a}
{Vanbeveren}, D., {Van Bever}, J., \& {Belkus}, H. 2007, \apjl, 662, L107

\bibitem[{{Vink} {et~al.}(2000){Vink}, {de Koter}, \& {Lamers}}]{Vink2000a}
{Vink}, J.~S., {de Koter}, A., \& {Lamers}, H.~J.~G.~L.~M. 2000, \aap, 362, 295

\bibitem[{{Vink} {et~al.}(2001){Vink}, {de Koter}, \& {Lamers}}]{Vink2001a}
{Vink}, J.~S., {de Koter}, A., \& {Lamers}, H.~J.~G.~L.~M. 2001, \aap, 369, 574

\bibitem[{{Walder} \& {Folini}(2000)}]{Walder2000a}
{Walder}, R. \& {Folini}, D. 2000, in Astronomical Society of the Pacific
  Conference Series, Vol. 204, Thermal and Ionization Aspects of Flows from Hot
  Stars, ed. H.~{Lamers} \& A.~{Sapar}, 281--+

\bibitem[{{Wheeler} {et~al.}(1987){Wheeler}, {Harkness}, {Barker}, {Cochran},
  \& {Wills}}]{Wheeler1987a}
{Wheeler}, J.~C., {Harkness}, R.~P., {Barker}, E.~S., {Cochran}, A.~L., \&
  {Wills}, D. 1987, \apjl, 313, L69

\bibitem[{{Woosley}(1993)}]{Woosley1993a}
{Woosley}, S.~E. 1993, \apj, 405, 273

\bibitem[{{Woosley} \& {Bloom}(2006)}]{Woosley2006a}
{Woosley}, S.~E. \& {Bloom}, J.~S. 2006, \araa, 44, 507

\bibitem[{{Woosley} \& {Heger}(2006)}]{Woosley2006b}
{Woosley}, S.~E. \& {Heger}, A. 2006, \apj, 637, 914

\bibitem[{{Yoon} {et~al.}(2006){Yoon}, {Langer}, \& {Norman}}]{Yoon2006a}
{Yoon}, S.-C., {Langer}, N., \& {Norman}, C. 2006, \aap, 460, 199

\end{thebibliography}

\end{document}